\documentclass[fleqn,usenatbib]{mnras}

\usepackage{newtxtext,newtxmath}
\usepackage[T1]{fontenc}
\usepackage{graphicx}
\usepackage{amsmath}
\usepackage{bm}
\usepackage{xcolor}
\usepackage{hyperref}
\usepackage{siunitx}
\usepackage{booktabs}
\usepackage{array}
\usepackage{multirow}
\usepackage{subcaption}
\usepackage{orcidlink}

\IfFileExists{floatpag.sty}{\usepackage{floatpag}\floatpagestyle{empty}}{}
\IfFileExists{lastpage.sty}%
  {\usepackage{lastpage}\newcommand{\lastpageno}{\pageref{LastPage}}}%
  {\newcommand{\lastpageno}{\pageref{lastpage}}}

\newcommand{\congr}{\texttt{CONGRuENTS}}

\newcommand{\abs}[1]{\left|#1\right|}

\newcommand{\pdv}[2]{\frac{\partial #1}{\partial #2}}
\newcommand{\msun}{M_{\odot}}
\newcommand{\kms}{\mathrm{km\ s^{-1}}}
\newcommand{\mh}{M_{\rm h}}
\newcommand{\mhn}{M_{\rm h,0}}
\newcommand{\mstar}{M_\star}
\newcommand{\mdotstar}{\dot{M}_\star}
\newcommand{\mcgm}{M_{\rm CGM}}
\newcommand{\mism}{M_{\rm ISM}}

\newcommand{\rvir}{r_{\rm vir}}
\newcommand{\rd}{r_{\rm d}}

\newcommand{\MA}{\mathcal{M}_{\rm A}}
\newcommand{\Mturb}{\mathcal{M}_{\rm turb}}
\newcommand{\Zfac}{\mathcal{Z}}

\newcommand{\Kdiff}{\mathcal{K}}

\newcommand{\aref}[1]{\hyperref[#1]{Appendix~\ref{#1}}}

\newcommand{\appropto}{\mathrel{\vcenter{
  \offinterlineskip\halign{\hfil$##$\cr
    \propto\cr\noalign{\kern2pt}\sim\cr\noalign{\kern-2pt}}}}}

\title[Cosmic rays in galaxy halos I]{Cosmic rays in galaxy halos I: the circumgalactic transport rate and a dynamically significant cosmic-ray reservoir}

\author[R. M. Crocker et al.]{
Roland M. Crocker,$^{\orcidlink{0000-0002-2036-2426},1}$\thanks{E-mail: rcrocker@fastmail.fm}
Mark R. Krumholz$^{\orcidlink{0000-0003-3893-854X},1}$
\\
$^{1}$Research School of Astronomy and Astrophysics, Australian National University, Canberra 2611, ACT, Australia
}

\date{Accepted XXX. Received YYY; in original form ZZZ}
\pubyear{2026}

\begin{document}
\label{firstpage}
\pagerange{\pageref{firstpage}--\lastpageno}
\maketitle

\begin{abstract}
\textit{Fermi}-LAT observations imply that most of the cosmic rays (CRs) produced in normal star-forming galaxies escape from the discs where they are accelerated into the circumgalactic medium (CGM), but the subsequent evolution of those escaping CRs is unclear. Here we construct a simple semi-analytic model for their transport and loss, which we constrain by confronting the model with a series of observations: limits on CR energy densities in the outer Milky Way from \textit{Fermi}-LAT, the isotropic $\gamma$-ray background (IGRB), stacked CGM X-ray emission from \textit{eRosita}, and the requirement that the accumulated pressure of CRs in galaxy CGMs not be so large as to render them unstable. 
We show that these constraints can be jointly satisfied in only two scenarios: one where CRs 
diffuse quickly through the CGM and freely escape beyond it
into the IGM, and another where the CR diffusion time out of halos is of order the Hubble time. 
In the latter scenario CRs accelerated by star formation naturally explain both the IGRB and the heat input required to maintain the temperature of the CGM against X-ray cooling, and their relatively slow transport means that roughly half of the CR energy injected into the CGM of an $L_\star$ halo over its star-formation history remains there at $z\simeq0$. 
This makes CR pressure dynamically significant, and also means that a non-negligible part of the IGRB is produced by CRs in the Milky Way's own halo. We argue that this latter scenario is the more likely one, as it avoids the need to posit a second source for the IGRB that is coincidentally very close in both power and spectral shape to the contribution from star-forming discs.
\end{abstract}


\section{Introduction}
\label{sec:intro}

Cosmic ray (CRs) accelerated by star formation are dynamically important in the interstellar media of star-forming galaxies \citep{Parker1966, Socrates2008, Zweibel2017, Krumholz2020, Crocker2021a, Ruszkowski2023}. A natural next question is:
{\it What about CRs in the circumgalactic medium (CGM)?}
Does the fraction of CR power that escapes galactic discs  accumulate in the CGM, modify the CGM pressure gradient, heat halo gas through streaming losses, or radiate hadronically at an observable level?
Here we show that the answer is, at least potentially: {\it yes}.

\subsection{Review of observational constraints}

This conclusion is inobvious, and is the subject of this series of papers, because constraints on CRs in the CGM are scarce and hard to interpret. Observationally, the most direct tracer of CR protons and heavier ions is the hadronic
(pion-decay) $\gamma$-ray emission they produce in collisions with ambient gas.
On disc scales, this emission shows that the densest starbursts are near-perfect calorimeters for CR protons, but that in normal star-forming galaxies only $\sim 10\%$ of CR protons interact before escaping into the CGM \citep{Thompson2007,Lacki2010,Strong2010,Mineo2012,Kornecki2020,Krumholz2020,Krumholz23a, Roth2021,Roth2023,Roth2024}.

However, their subsequent fate has proven difficult to determine. Once CRs escape from the dense interstellar medium (ISM) and reach the much more diffuse CGM, observing them becomes difficult because the expected surface brightness is low and
the foregrounds severe. Measurements of the $\gamma$-ray emissivity of gas in the outer Milky Way provide some limited constraints \citep{Acero2016,Yang2016,Pothast2018}, but thus far the only claimed detections of extragalactic CGM $\gamma$-ray emission are for M31 
\citep{Karwin2019,Recchia2021,Roy2022b} and from stacking \citep{Pshirkov2026}. However, the uncertainties are very large and even the source of emission (e.g., hadronic versus leptonic) is uncertain. Recent cosmological simulations that follow CR transport predict extended $\gamma$-ray --- and associated soft X-ray
inverse-Compton --- halos at potentially detectable levels \citep{Hopkins2025,Ponnada2025,Ponnada2026,Lu2026}. However, these methods are necessarily degenerate: the hadronic yield is set by the \emph{product} of the CR and gas densities, so halo $\gamma$-rays constrain that product and not either factor by itself. Consequently, it has not yet proven possible to draw general conclusions from these limited data. 

The fate of CRs in the CGM is also constrained by the isotropic $\gamma$-ray background \citep[IGRB;][]{Ackermann2015}, which bounds both emission from the Milky Way's own CGM and the collective output of CR protons in external galaxies, whether it arises in discs or halos. For the former, the possibility that the MW CGM contributes appreciably has been raised \citep{Feldmann2013}, and \citet{Liu2019} show that the IGRB already bounds the Galactic CR luminosity and the halo baryon budget, depending on the diffusion coefficient. For the latter, \citet{Roth2021} show that galaxy discs and their CGMs cannot be full proton calorimeters because, if they were, the total $\gamma$-ray output of star-forming galaxies would exceed the IGRB \citep[see the $f_{\rm cal} = 1$ curve of fig.~4 of][]{Roth2021}; a significant fraction of the CR protons accelerated over cosmic history must instead lose their energy non-radiatively (e.g., via streaming) or survive to the present. As for individual halos, however, conclusions stronger than upper limits are troubled by the same degeneracy: the IGRB constrains only the product of the CR proton and CGM gas densities.

CR electrons offer further constraints in principle. Radio synchrotron halos around star-forming galaxies trace the CR-electron population and its transport out of the disc \citep{Beck2015,Heesen2018,Heesen2023}, and Faraday rotation of background sources probes the CGM magnetic fields on which those electrons radiate \citep{Bernet2008,Bernet2010,Mao2010}, partially breaking the electron--magnetic-field degeneracy that is the analogue of the proton--gas degeneracy above. Radio halos and relics, rotation measures and $\gamma$-ray limits have long constrained CR populations and magnetic fields in cluster atmospheres \citep{Carilli2002, Ensslin2011, Brunetti2014}, but applying these techniques to the far dimmer emission of lower-mass halos is in its infancy. In soft X-rays, eROSITA detects emission around $L_\star$ halos extending further than purely thermal models predict \citep{Predehl2021eROSITA,Zhang2024profiles,Zhang2024eROSITAII}, prompting the suggestion that inverse-Compton emission from CR electrons contributes \citep{Hopkins2025,Lu2026}; however, this hinges on the claim that the emission cannot be thermal, in turn dependent on uncertain models of galactic winds.

Independently, UV and X-ray absorption-line studies of the cool and warm-hot CGM also constrain CRs indirectly, through their contribution to the total pressure (\citealt{Werk2014}; the O\,\textsc{viii} analysis of \citealt{Roy2022}). Phenomenological models calibrated against COS-Halos O\,\textsc{vi} and Milky Way O\,\textsc{vii} and O\,\textsc{viii} absorption and emission suggest that galaxy CGMs contain massive, comparatively long-lived reservoirs of hot gas supported roughly equally by thermal pressure, turbulence, and magnetic and CR pressure \citep{Faerman2017,Faerman2020,Faerman2022} --- an inference from the gas ionisation state that is independent of the route we pursue here.

\subsection{Review of theoretical models}

Theoretically, the case that CRs matter in the CGM is strong but model-dependent.
Galaxy-formation simulations that follow CR transport find that CRs can come to
dominate the pressure in the inner CGM, flatten its gas density profile, and drive
or reshape galactic winds
\citep{Ji2020, Butsky2023, Hopkins2021, Pakmor2020, Yamasaki2024};
see \citet{Ruszkowski2023} for a review. However, these outcomes depend sensitively on the
assumed CR transport --- whether streaming- or diffusion-dominated, and on the
effective diffusivity --- which remains uncertain at the order-of-magnitude level
and is set by poorly-constrained micro-scale plasma physics
\citep{Zweibel2017, Hopkins2022, Reichherzer2025, Habegger2026}.

Pinning this transport
down empirically, across the range of halos that contribute to the integrated
signals, is precisely what the halo-scale $\gamma$-ray data can do --- provided
the comparison is made in a framework that respects the time-dependence of the
problem. CR transport and loss times 
in the CGM are often comparable to halo growth times and to cosmological lookback times,
so the CR hadron population inherited by a halo at low redshift depends on the coupled evolution of halo mass, host-galaxy star formation, halo gas structure, and CR transport, and cannot be assumed to be in steady state. Moreover, the uncertainty about CR transport is also tied to the question of how gas is arranged \textit{within} the CGM: CRs can alter this distribution via their pressure, but at the same time the gas distribution strongly affects the observable emission, with models in which CR transport is slow and CGM gas is highly concentrated toward central galaxies yielding vastly higher emission than those where either CR transport is faster or CGM gas less concentrated.

\subsection{Goals and plan for this series of papers}

Here we set out a mathematically coherent treatment of CR transport and energy loss in the halos of galaxies that is capable of generating predictions that we can compare to  the observational and dynamical constraints we have discussed. We focus on CR protons, which dominate the energy budget, and for brevity from this point forward we will drop the `proton' qualifier, so the generic term CR refers to CR protons unless otherwise specified.
Our model combines (i) an evolving halo-plus-galaxy framework, (ii) a host-galaxy CR injection and escape model informed by \congr{} \citep{Roth2021,Roth2023,Roth2024}, and (iii) a redshift-dependent transport equation for CRs propagating through a cored CGM gas profile under several boundary-condition assumptions that span the range of physically plausible scenarios. 
Our implementation treats macroscopic CR transport and halo gas distributions  via parametrised functions agnostic as to the detailed physics underlying them, allowing us to marginalise over these uncertainties while still asking the question:
{\it does there exist a model that satisfies the totality of empirical constraints?}

This paper (Paper I) is the first of a series of three. It describes our model and its immediate implications for the energetics of galaxy halos and $\gamma$-ray budgets, adopting empirically motivated fiducial parameters. Paper II develops the multi-messenger signals generated by the resulting population of CRs, and Paper III the implications of our findings for the physical model of CR transport in the CGM, showing that the totality of the observational constraints sets limits on the transport rate and hints at particular classes of transport model.

The remainder of this paper is organised as follows. In \autoref{sec:constraints} we detail the empirical constraints we use to decide which models are viable, and why; in \autoref{sec:model} we introduce our model for CRs in the CGM, including halo growth and star formation, CR injection, and CR transport and loss; and in \autoref{sec:results} we confront models with the observations to determine what range of models matches them. We discuss the implications of the observational constraints in \autoref{sec:discussion}, and summarise and conclude in \autoref{sec:conclusion}.

\section{Empirical constraints}
\label{sec:constraints}

We begin by cataloguing the empirical constraints we use to evaluate our models: the IGRB (\autoref{ssec:IGRB}), the X-ray luminosity of $L_\star$-galaxy CGMs (\autoref{ssec:erosita}), the $\gamma$-ray luminosity of the outer Milky Way disc (\autoref{ssec:mw_profile}), the pressure balance of CGMs (\autoref{ssec:pressure_support}), and the stability of halo virial shocks (\autoref{ssec:confinement}). \autoref{ssec:combined_constraints} gives a qualitative picture of how they fit together and why their combination pins down the fate of CRs in the CGM.

\subsection{The IGRB}
\label{ssec:IGRB}
\label{app:IGRB_datum}

As discussed in \autoref{sec:intro}, the brightness of the IGRB sets one important constraint on the fate of CRs: if all the CRs produced by SFRs over cosmological star formation history were to lose their energy to {\it pp} collisions, or if too many CRs were to survive in the MW CGM at the present day, the resulting $\gamma$-ray luminosity would significantly exceed the observed IGRB.

Quantitatively, we adopt the model-A IGRB data of \citet{Ackermann2015}, and from these data we reconstruct the integrated all-sky IGRB energy flux in the \(1\text{--}10~{\rm GeV}\) band, including the partially overlapping edge bands at \(1\) and \(10~{\rm GeV}\). We focus on the \(1\text{--}10~{\rm GeV}\) range because these energies are largely unaffected by $\gamma\gamma$-opacity effects, thereby removing any dependence of our results on a choice of model for the extragalactic background light. We first convert \citeauthor{Ackermann2015}'s tabulated photon intensities to energy fluxes assuming a within-band power-law spectrum with photon index \(\Gamma=2.32\), consistent with the model-A spectral fit of \citet{Ackermann2015}, and then sum the per-band fluxes over the \(1\text{--}10~{\rm GeV}\) interval. This gives a central estimate $F_{\rm central}^{\rm IGRB}(1\text{--}10~{\rm GeV}) = 1.38\times10^{-8}\,{\rm erg\,cm^{-2}\,s^{-1}}$.
Propagating in quadrature the tabulated model-A statistical/instrument uncertainties across the contributing bands gives $\sigma_{\rm stat}^{\rm IGRB} = 6.78\times10^{-10}\,{\rm erg\,cm^{-2}\,s^{-1}}$, and doing the same for the tabulated foreground-modelling uncertainties gives an asymmetric systematic uncertainty $\sigma_{{\rm sys},\pm}^{\rm IGRB} = (-2.15,+2.35)\times10^{-9}\,{\rm erg\,cm^{-2}\,s^{-1}}$; for simplicity we ignore the fact that foreground modelling errors are somewhat correlated across bands, an effect that if included would somewhat broaden the uncertainties. Because the systematic uncertainty is asymmetric and does not naturally behave as a Gaussian ``\(1\sigma\)'' error, we do not fold it into a single effective variance when comparing against observations. Instead, for an upper-limit style exclusion test we define an allowed flux
\begin{eqnarray}
F_{\rm allow}^{\rm IGRB} &=& F_{\rm central}^{\rm IGRB} + 3\,\sigma_{\rm stat}^{\rm IGRB} + \sigma_{{\rm sys},+}^{\rm IGRB} \nonumber \\
&=& 1.82\times10^{-8}\,{\rm erg\,cm^{-2}\,s^{-1}},
\label{eq:igrb_1_10_allow}
\end{eqnarray}
all for the \(1\text{--}10~{\rm GeV}\) band. Models for which the predicted integral \(1\text{--}10~{\rm GeV}\) flux exceeds \(F_{\rm allow}^{\rm IGRB}\) are considered excluded by IGRB overproduction.

In addition to this exclusion test, we also assess whether models are \textit{sufficient} to explain the IGRB by evaluating whether their predicted IGRB luminosity (again summing the contributions from the Milky Way's own CGM, extragalactic star-forming discs, and extragalactic CGMs) reaches the $1\sigma$ lower edge of the observed IGRB. Quantitatively, the flux required to meet this requirement is $F_{\rm suff}^{\rm IGRB}=F_{\rm central}^{\rm IGRB}-\bigl[(\sigma_{\rm stat}^{\rm IGRB})^2+(\sigma_{{\rm sys},+}^{\rm IGRB})^2\bigr]^{1/2}$, which evaluates to $F_{\rm suff}^{\rm IGRB}=1.14\times10^{-8}\,{\rm erg\,cm^{-2}\,s^{-1}}\simeq0.82\,F_{\rm central}^{\rm IGRB}$.
Note that in evaluating \(F_{\rm suff}^{\rm IGRB}\)
we adopt the larger (upward) foreground-modelling uncertainty, \(\sigma_{{\rm sys},+}^{\rm IGRB}\), rather than \(\sigma_{{\rm sys},-}^{\rm IGRB}\). This lowers the threshold and so admits more of the parameter space; with \(\sigma_{{\rm sys},-}^{\rm IGRB}\), \(F_{\rm suff}^{\rm IGRB}\) would rise to \(0.84\,F_{\rm central}^{\rm IGRB}\).
While we assess whether models are sufficient to produce the full IGRB, we do not exclude those that are insufficient, because there is considerable debate in the literature about the relative contribution of star-forming galaxies versus other sources (largely blazars) to the IGRB \citep[e.g.,][]{Zechlin2016a,Manconi2020,Eckner2025}.

\subsection{CGM X-ray luminosities}
\label{ssec:erosita}

If CRs that enter the CGM dissipate their energy via non-radiative loss mechanisms, most prominently the streaming instability, this power must eventually emerge as thermal radiation from CGM gas. The observed luminosity of this gas then sets an upper limit on the rate at which CRs can dissipate energy non-radiatively. 
We derive our upper limit from the eROSITA stacking analysis of \citet{Zhang2024eROSITAII}, who report the mean rest-frame \(0.5\)–\(2\) keV luminosity of the CGM, \(L_{X,\rm CGM}\), within a radius inside which the mean density is 500 times the critical density (compared with
200 times the \emph{mean matter} density, which defines the virial radius and halo mass throughout
this paper; \autoref{ssec:halos}).
They report this value as a function of the halo mass within this radius over a range $\log (M_\mathrm{500c}/\mathrm{M}_\odot) = 11.3 - 13.7$. For the purposes of our analysis, we interpolate their tabulated values of $L_{X,\rm CGM}$ as a function of $\log M_\mathrm{500c}$, together with the uncertainty on the stacked mean luminosity converted from linear to logarithmic coordinates, to generate a predicted X-ray luminosity as a function of present-day halo mass $\log\mh$, self-consistently accounting for the difference between $\log\mh$ and $\log M_{\rm 500c}$ based on our adopted halo density profile and concentration parameter (see \aref{app:cgm_binding_energy}). For a present-day halo mass $\mh = 10^{12}$ M$_\odot$ we have $M_{\rm 500c} = 0.61\mh$, and our interpolation procedure gives $L_{X,\rm CGM} = 5.5\times 10^{39}$ erg s$^{-1}$ with an uncertainty of $\sigma_{X,\rm CGM} = 0.208$ dex. 

In analogy with our treatment of the IGRB, we consider a model to be excluded by the observed CGM luminosity if it predicts that the CGM X-ray luminosity $L_{X,\rm CGM} > L_{X,\rm CGM,allow}$, where $\log L_{\rm X,CGM, allow} = \log L_{X,\rm CGM} + 2\sigma_{X,\rm CGM}$,
with $L_{X,\rm CGM}$ and $\sigma_{X,\rm CGM}$ taken from our fit to \citet{Zhang2024eROSITAII}. For a Milky Way-mass halo, $\mh = 10^{12}$ M$_\odot$, this evaluates numerically to 
$L_{X,\rm CGM, allow} = 1.5\times 10^{40}$ erg s$^{-1}$. We may also adopt the hypothesis that the CGM is in fact heated primarily by the non-radiative dissipation of CR energy, in which case we expect the CGM luminosity to be approximately equal to the rate of non-radiative CR losses, rather than an upper limit to it. In analogy with our treatment of the IGRB, we therefore consider non-radiative losses sufficient to explain the observed X-ray emission from the CGM for a given model if the predicted X-ray luminosity is within the range $\log L_{X,\rm CGM,suff} = \log L_{X,\rm CGM} \pm \sigma_{X,\rm CGM}$, which evaluates numerically to $L_{X,\rm CGM,suff} = 5.5_{-2.1}^{+3.4} \times 10^{39}$ erg s$^{-1}$.

This scenario for CGM X-ray emission is quite distinct from the model of \citet{Hopkins2025}, in which CGM X-rays are inverse-Compton emission from CR leptons. Although both invoke CRs as the ultimate power source, they predict different spectra: a line-free, non-thermal spectrum in the \citeauthor{Hopkins2025} model, but here the normal complement of lines from a thermally emitting plasma, since CRs dissipate their power into heat.

\subsection{$\gamma$-ray emission from the outer Milky Way}
\label{ssec:mw_profile}

The $\gamma$-ray emissivity of hydrogen in the outer Milky Way disc is constrained by \textit{Fermi}-LAT measurements of $\gamma$-ray emission combined with 21 cm maps of neutral hydrogen emission, where the velocity can be mapped to a kinematic distance assuming a disc rotation curve \citep{Acero2016,Yang2016,Pothast2018}. Since these data constrain the emissivity per H nucleon, they set an upper limit on the CGM CR energy density in the vicinity of this gas -- an upper limit only, since most of the emission from this gas likely arises from a disc component of CRs rather than from CRs in the CGM interacting with gas in the disc. We take our estimates of the CGM CR energy density to be $u_c <  0.37\pm0.14\,$eV\,cm$^{-3}$ from 1--10\,GeV and $u_c < 0.12\pm0.04\,$eV\,cm$^{-3}$ above 10\,GeV at $r\simeq 8$ kpc \citep{Yang2016}. We place our maximum allowed energy densities at the $2\sigma$ upper limits, i.e., we consider a model to be excluded if it predicts $u_c > u_{c,\rm allow}$, where $u_{c,\rm allow} = 0.65$ eV cm$^{-3}$ from $1-10$ GeV and 0.20 eV cm$^{-3}$ at $>10$ eV. Unlike with the CGM and IGRB constraints, we do not define a corresponding sufficiency condition, since in practice we do not expect $u_c$ to be dominated by CGM CRs, but by CRs produced locally in the disc.

Our approach to these data differs considerably from that of \citet{Hopkins2022, Hopkins2025}, who treat the observed emission as a constraint on the CGM surface brightness at a projected radius equal to the kinematically inferred disc radius. This implicitly treats $\gamma$-rays from disc CRs interacting with disc gas as quasi-spherical emission from CGM CRs interacting with CGM gas, which seems questionable, particularly because the inferred CR spectrum is very soft \citep{Pothast2018}, as expected for a disc population and inconsistent with the harder spectrum expected for CRs in the CGM. Treating these data as upper limits only is a far more conservative --- and in our view more defensible --- approach.

\subsection{CGM pressure balance}
\label{ssec:pressure_support}

Another constraint is that CR pressure in galactic CGMs cannot grow so large that it prevents gas from achieving hydrostatic balance, at least in halos massive enough that we expect the CGM to be quasi-hydrostatic. Specifically, if the CR energy density (integrated over all CR energies) is $u_c$, there is a corresponding pressure $P_c = u_c/3$ (assuming the CRs are predominantly relativistic), and we can define the fraction of the hydrostatic pressure provided by CRs as
\begin{equation}
        \eta_P = \left|\frac{\partial P_c/\partial r}{\rho_g g}\right|,
    \label{eq:etaP_def}
\end{equation}
where $r$ is the radius within a halo, $\rho_g$ is the gas density, and $g$ is the gravitational acceleration. We defer a discussion of how we compute $\eta_P$ in practice from our model to the next section, but we consider any model for which the predicted value of $\eta_P$ exceeds unity for an $L_\star$ halo ($M_{\rm h,0}=10^{12}\,\msun$) at $z=0$ to be excluded on the grounds that the CR energy density would be too high to permit the existence of a hydrostatic CGM. Note that this is a generous choice, since $\eta_P = 1$ corresponds to the CGM being supported \textit{entirely} by CR pressure, with zero contribution from thermal, magnetic, or turbulent pressure.

\subsection{Virial shock confinement}
\label{ssec:confinement}

In addition to the pressure gradient of CRs \textit{within halos}, in models where CRs are assumed to be confined by a virial shock (cf.~\autoref{sec:model}), we  require that the CR pressure at the virial radius not exceed the ram pressure of cosmological infall onto the halo, $P_{\rm c}(r_{\rm vir})\le P_{\rm ram}$. We defer a discussion of the numerical implementation of this constraint in the context of our model to \autoref{ssec:model_constraints}.

\subsection{Combined constraints}
\label{ssec:combined_constraints}

The constraints we have outlined combine usefully because they bound the problem from complementary directions. That galaxy \textit{discs} are not perfect CR calorimeters sets a lower limit on the rate at which CRs enter the CGM; the question then becomes how they exit. There are four possibilities: escape into the intergalactic medium (IGM), loss via radiative mechanisms, loss via non-radiative mechanisms, and no loss at all, i.e., the CRs that entered the CGM over cosmic time are still there today.

Each possibility is constrained by one or more of our conditions. The IGRB (\autoref{ssec:IGRB}) limits the fraction of CRs entering the CGM that can dissipate their energy in {\it pp} collisions, which produce observable $\gamma$-rays. Non-radiative losses that heat the CGM are constrained by CGM X-ray luminosities (\autoref{ssec:erosita}). The possibility that CRs have simply accumulated and are still resident is limited both by the $\gamma$-ray emissivity of gas in the outer Milky Way disc (\autoref{ssec:mw_profile}) and by the requirement that the accumulated pressure neither disrupt the CGM (\autoref{ssec:pressure_support}) nor prevent formation of an assumed virial shock (\autoref{ssec:confinement}). We will see that the combination of all these conditions eliminates a significant portion of the parameter space for CR transport and yields circumstantial evidence for a specific scenario.

\section{Model}
\label{sec:model}

Here we describe our model for CRs in the CGM, which we have built to take advantage of the constraining power offered by the observations summarised in \autoref{sec:constraints}. In our model we consider a spherically-symmetric CGM through which CRs propagate and with which they interact to produce observable emission. To develop this model, we first introduce a treatment of cosmological halos and their baryonic -- stellar, ISM, and CGM -- contents, which serves as a backdrop for our CR transport model (\autoref{ssec:halos}). We next introduce a model for CR boundary conditions: the rate at which CRs are injected by the central galaxy, and what happens when they reach the edge of the halo (\autoref{ssec:CR_boundaries}). With the background and boundary conditions established, we then introduce our CR transport equation and our numerical method for solving it (\autoref{ssec:transport}). Finally, we explain how we go from a solution to the model a prediction for each of the empirical constraints (\autoref{ssec:model_constraints}).

\subsection{Cosmological halo evolution}
\label{ssec:halos}

Here and throughout this work we adopt a flat Planck cosmology \citep{Planck2020} with present-day matter and cosmological constant densities $\Omega_{\rm m, 0} = 0.315$ and $\Omega_{\Lambda,0} = 0.685$, baryon fraction $f_{\rm b} = 0.157$, spectral index $n_s = 0.965$, density variance $\sigma_8 = 0.8$, CMB temperature $T_\mathrm{CMB} = 2.7255$ K, and present-day Hubble constant $H_0 = 100h\,\kms\,\mathrm{Mpc}^{-1}$ with  $h=0.674$. For this cosmology the present-day critical density is $\rho_\mathrm{crit,0} = 2.78\times 10^{11}h^2\msun\,\mathrm{Mpc}^{-3}$, the mean cosmological matter density evolves as $\rho_{\rm m}(z) = \Omega_{\rm m,0}\rho_\mathrm{crit,0} (1+z)^3$, and the Hubble parameter evolves as $H(z) = H_0 \sqrt{\Omega_{\rm m,0}(1+z)^3+\Omega_{\Lambda,0}}.$ Cosmic time and redshift are related via the usual $dt/dz = -[(1+z)H(z)]^{-1}$.

\subsubsection{The dark matter component}

Within this cosmology we follow the evolution of a halo of present-day dark matter mass $\mhn$ using a Neistein-Dekel style growth formalism \citep{Neistein2008, Bouche2010}, whereby the time evolution of the halo mass is given by
\begin{equation}
\frac{d M_{\rm h,12}}{dt} = -0.628\,M_{\rm h,12}^{1.14}\,\dot\omega(z),
\label{eq:mh12dot_backbone}
\end{equation}
where $M_{\rm h,12} = M_{\rm h}/(10^{12}\,\msun)$ and
$\dot\omega(z) = -0.0476 [1\ + \ z\ +\ 0.093(1+z)^{-1.22}]^{2.5}\,{\rm Gyr}^{-1}.$
We generate tracks of $\mh(z)$, the mass as a function of redshift for a halo of present-day mass $\mhn$, by numerical integration of \autoref{eq:mh12dot_backbone}. The virial radius is
\begin{equation}
    r_\mathrm{vir}(z;\mhn) = \left[\frac{3\mh(z)}{4\pi\,200\,\rho_{\rm m}(z)}\right]^{1/3},
\label{eq:rvir_exact}
\end{equation}
We limit our calculations to halos with present-day masses $10^{10} \leq M_{\rm h,0}/\msun \leq 10^{13}$, a range over which this approximation is reasonable, and which dominates cosmological star formation.

To generate predictions for cosmologically-integrated quantities such as the IGRB, we also require a halo mass function. We follow \citet{Tinker2008} by adopting a comoving halo number density
\begin{equation}
    \frac{dn}{d\mh}(\mh,z) = f(\sigma)\,
    \frac{\rho_{\rm m}(0)}{\mh}\,
    \frac{d\ln \sigma^{-1}}{d\mh},
    \label{eq:halo_density_tinker}
\end{equation}
where $\rho_{\rm m}(0)$ is the present-day comoving matter density, $\sigma(\mh,z)$ is the variance of the linear density field smoothed on the scale enclosing mass $\mh$, and \(f(\sigma)=A\left[(\sigma/b)^{-a}+1\right]\exp(-c/\sigma^{2})\) is an empirical fitting function. We adopt the coefficient set \((A,a,b,c)=(0.186,\,1.47,\,2.57,\,1.19)\), corresponding to \citeauthor{Tinker2008}'s \(\Delta=200\) spherical-overdensity case. The density variance is separable to good accuracy over the redshift range of interest, \(\sigma(\mh,z)=\sigma(\mh,0)\,D(z)\), with \(D(z)\) the linear growth factor normalised to \(D(0)=1\). We evaluate \(\sigma(\mh,0)\) directly from the linear matter power spectrum \(P(k)\propto k^{n_s}T^{2}(k)\), smoothed with a top-hat window on the scale \(r\) enclosing the mass, \(\mh=(4\pi/3)\rho_{\rm m}(0)r^{3}\), where \(T(k)\) is the zero-baryon transfer function of \citet{EisensteinHu1998}, which is accurate to \(\simeq2\%\) for our cosmology. The normalisation is fixed by requiring \(\sigma=\sigma_8\) at \(r=8\,h^{-1}\,{\rm Mpc}\). 

\subsubsection{The baryonic disc component}

We assign stellar masses and star formation rates to halos
using a slightly-modified version of the redshift-dependent stellar-to-halo mapping from \citet{Behroozi2013}. Specifically, we take the stellar mass for a halo of mass $\mh$ at redshift $z$ to be given by
\begin{equation}
    \log\mstar = \log(\epsilon M_1)
    + f_{\rm B13}\left(\log\frac{\mh}{M_1}\right) - f_{\rm B13}(0) + \log M_\mathrm{reg}
\label{eq:mstar_mh}
\end{equation}
and the corresponding star formation rate to be
$ \dot{M} = 1.35\ d\mstar/dt,$
where the prefactor approximately accounts for stellar mass returned to the ISM \citep{van_Dokkum2013}. Here
\begin{equation}
f_{\rm B13}(x)
= -\log(10^{\alpha x}+1) + \delta\,
(1+e^{10^{-x}})^{-1}
[\log(1+e^x)]^{\gamma},
\end{equation}
and the quantities $\epsilon(z)$, $M_1(z)$, $\alpha(z)$, $\delta(z)$, and $\gamma(z)$ are redshift-dependent fitting functions that we take directly from the code implementation of \citeauthor{Behroozi2013} The final term $M_\mathrm{reg}$ in \autoref{eq:mstar_mh} is a regularising term that we add to the \citeauthor{Behroozi2013} formalism:
\begin{equation}
M_\mathrm{reg} = 10^{8.8} e^{-z/0.070}\, 
[1/2 \tanh(\log\mh-12.4)+1]\ \msun.
\label{eq:b13_regulariser}
\end{equation}
This term makes almost no difference to $\mstar$ itself, but has the important effect of ensuring that $d\mstar/dt$ remains positive-definite, whereas in the unmodified \citeauthor{Behroozi2013} model it becomes slightly negative for halos with masses $\gtrsim 10^{12}\,\msun$ at $z \lesssim 0.05$, giving a physically-impossible negative star formation rate. We choose our functional form for $M_\mathrm{reg}$ purely empirically so the star formation rates we obtain for $\mhn \sim 10^{12}\msun$ halos and the total cosmological star formation rate density at $z=0$ match observations reasonably well, something we demonstrate below.

In our model we will also require an estimate for size and gas mass of the central galaxy. Following \citet{Forbes2014} and \citet{Krumholz2018}, we adopt a disc radius $\rd = R_0 \rvir$ with $R_0 = 0.018$.\footnote{Here and throughout, we will use lowercase $r$ to indicate physical radii, and uppercase $R$ to indicate dimensionless radii normalised to $\rvir$.}
We then infer the ISM gas mass from the inverted Kennicutt--Schmidt relation as implemented by \citet{Roth2021,Roth2023}:
\begin{equation}
\mism
\simeq 1.91\times 10^{10}\,\msun
\left(\frac{\mdotstar}{\msun\,\mathrm{yr}^{-1}}\right)
\left(\frac{\mstar/\msun}{2\pi \rd^2/\mathrm{pc}^2}\right)^{-0.48}.
\label{eq:mism_backbone}
\end{equation}

\subsubsection{The CGM}
\label{sssec:cgm_model}

Finally, we assign all the remaining baryonic mass to the CGM,
\begin{equation}
\mcgm=f_{\rm b}\,\mh-\mstar-\mism.
\label{eq:mcgm_backbone}
\end{equation}
We assume that within the virial radius this gas is distributed in a cored powerlaw profile
\begin{equation}
    \rho_{\rm g,in}(R) = \rho_0\,(R^2+R_0^2)^{-\alpha/2},
    \label{eq:cored_profile}
\end{equation}
for $R \in (R_0,1)$, where $R=r/\rvir$ is a dimensionless radius and the index $\alpha$ is left as a free parameter. This index must be the range $1 < \alpha < 3$ for the enclosed mass and column density to be well-behaved; the common isothermal assumption corresponds to $\alpha = 2$.

The choice of normalisation $\rho_0$ is subtle because it is tied to the question of how gas near the virial radius is structured, a topic on which the literature has not yet converged. The spherical stability analysis of \citet{Birnboim2003}, extended by \citet{Dekel2006}, predicts that only halos with mass $\mh\gtrsim10^{12}\,\msun$ are bounded by virial shocks (at $z=0$), while some simulations find a shock-bounded hot halo already above $\sim10^{11.5}\,\msun$ \citep{Fielding2017} and others find that the inner CGM virialises only once $\mh$ exceeds $\sim10^{12}\,\msun$ \citep{Stern2021}, and even then that the outer CGM develops a non-hydrostatic configuration rather than a sharp spherical shell \citep{Lochhaas2023}. The presence or absence of a shock can also be affected by CRs themselves: \citet{Ji2021} find that in Milky Way-mass halos at $z\lesssim1$ the virial shock is suppressed altogether once CR pressure dominates the thermal pressure because infalling gas is decelerated gradually by the CR pressure gradient and remains subsonic out to and beyond $\rvir$. 
Their runs adopt $\kappa\simeq3\times10^{29}\,{\rm cm^{2}\,s^{-1}}$, within the range we infer below (\autoref{subsec:parameter_space_scans}), so the character of the outer boundary is itself partly contingent on the CR content we set out to measure. 

For this reason, we do not adopt a preferred configuration, and instead consider three distinct scenarios for the CGM density profile $\rho_\mathrm{g,out}$ outside $R=1$, and for the closure radius $R_\mathrm{clos}$, defined implicitly as the radius for which 
\begin{equation}
    4\pi \rvir^3 \int_{R_0}^{R_\mathrm{clos}} \rho_{\rm g} R^2 \, dR = \mcgm.
    \label{eq:Rclose_integral}
\end{equation}
We name each scenario after the cosmic-ray outer boundary it implies (\autoref{ssec:CR_boundaries}). Our scenarios are:
\begin{itemize}
    \item \textsc{TrappingMax}: this is the simplest case, in which we assume that there is a virial shock present, and that this shock is able to confine the full baryonic mass of the halo to within the virial radius. In this case we adopt $R_\mathrm{clos} = 1$ and $\rho_{\rm g,out} = 0$, which in turn fixes the value of $\rho_0$. We refer to this as the \textsc{TrappingMax} case because it maximises the density interior to $\rvir$.
    \item \textsc{TrappingMin}: in this case we assume that a virial shock exists, but that some of the CGM mass is outside it so that $R_\mathrm{clos} > 1$. The mass outside $R = 1$ has a powerlaw density profile similar to that of the interior mass, $\rho_\mathrm{g,out} = \rho_{\rm vir} R^{-\alpha}$, and the inner and outer density normalisations $\rho_0$ and $\rho_{\rm vir}$ are related implicitly by requiring that $\rho_{\rm g,in}(1) = \rho_\mathrm{vir}/4$, i.e., there is a factor of four density jump across the virial shock, as expected for a strong shock in $\gamma=5/3$ adiabatic gas. We then minimise the density interior to $R=1$ by taking $R_\mathrm{clos}$ to be the radius at which the density falls to the mean matter density of the Universe, i.e., we require $\rho_{\rm g,out}(R_\mathrm{clos}) = \rho_\mathrm{m}$. This condition together with the integral requirement given by \autoref{eq:Rclose_integral} constitute two equations in the two unknowns $R_\mathrm{clos}$ and $\rho_{\rm vir}$, which we can solve numerically and thereby obtain the full density profile. The fraction $f_\mathrm{in}$ of the CGM mass interior to the virial shock, and thus reduction in the value of $\rho_0$ for this scenario compared to \textsc{TrappingMax}, depends on the choice of $\alpha$ and (very weakly) on the halo mass -- for $\alpha = 2$, we find numerically that $f_\mathrm{in}\approx 0.65$, and that $f_\mathrm{in}$ ranges from $\approx 0.15 - 0.95$ as we vary $\alpha$ from 1 to 3.
    \item \textsc{FreeEscape}: in this scenario we assume that there is no virial shock, and thus the inner density profile given by \autoref{eq:cored_profile} continues unchanged past $R=1$. We then set the value of $R_\mathrm{clos}$ to that at which the density falls to ten times the mean matter density of the Universe, i.e., $\rho_\mathrm{g}(R_\mathrm{clos}) = 10 \rho_{\rm m}$. This condition in turn requires $R_\mathrm{clos} = \max[1, (20x/3)^{1/\alpha}]$, where $x = 1 - (\mstar+\mism)/f_b \mh$, and using this value of $R_\mathrm{clos}$ in \autoref{eq:Rclose_integral} then determines $\rho_0$.
\end{itemize}

\subsubsection{Numerical validation}

\begin{figure}
\includegraphics[width=\columnwidth]{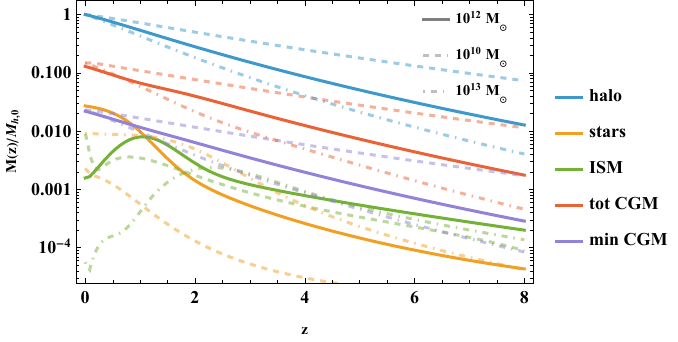}
\caption{Redshift evolution of the masses of the various halo components within our model normalised to the present-day total mass of a halo, $M_{\rm h,0}$, for three different present-day halo masses. The two lines ``tot CGM'' and ``min CGM'' correspond to the total $\mcgm$ (\autoref{eq:mcgm_backbone}) and to the mass inside $\rvir$ in the \textsc{TrappingMin} scenario with $\alpha\to 1$, respectively.
}
\label{fig:plotMassEvolution}
\end{figure}

To validate that our prescriptions produce reasonable outcomes, and give a qualitative overview of what they predict, in \autoref{fig:plotMassEvolution} we plot the time evolution of the halos mass, stellar mass, ISM mass, and CGM mass (both total and only that inside $\rvir$) for present-day halo masses $\mhn = 10^{10}, 10^{12}$ and $10^{13}\,\msun$. We see that the outcomes are qualitatively reasonable: the present-day stellar mass for a Milky Way-like halo, $\mhn = 10^{12}\,\msun$, is a few times $10^{10}\,\msun$, and the present-day ISM mass is $\approx 10\%$ of this. The ISM mass reaches near-equality with the stellar mass at $z\sim 2$. More massive halos are very ISM-poor at $z=0$, and reach ISM-stellar equality at higher $z$, while the $\mhn=10^{10}\,\msun$ remains gas-dominated even at $z=0$.

\begin{figure}
    \includegraphics[width=0.48\textwidth]{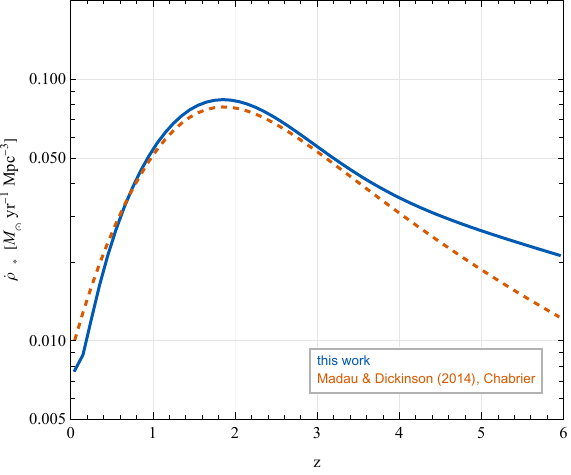}
    \caption{The cosmic star-formation history of the Universe predicted by our model (solid blue line) compared to the observed cosmic star formation history \citet{Madau2014} converted to a \citet{Chabrier2005} initial mass function (dashed orange line).}
    \label{fig:cosmicvalidation}
\end{figure}

We can also verify that our prescriptions produce reasonable outcomes for the integrated star formation rate density of the Universe:
\begin{equation}
    \dot{\rho}_\star(z) = \int \frac{dn}{d\mh} \mdotstar(\mh,z) \, d\mh.
\end{equation}
We evaluate this integral over the fiducial range covered by our models, $10^{10} \leq \mhn/\msun \leq 10^{13}$, and plot the results in \autoref{fig:cosmicvalidation}. For comparison, we also show the cosmological star formation rate density estimated by \citet{Madau2014} converted to a \citet{Chabrier2005} IMF. The figure shows that our simple analytic model agrees with the observations to within $\approx 15\%$ over $0.5 < z < 3.5$, which is the range of greatest importance for our calculations. The $\sim 20\%$ disagreement between the true star formation rate and that predicted by our model at $z = 0$ is likely due to our omission of halos smaller than $10^{10}\,\msun$, while the divergence at $z\gtrsim 5$ is likely because the fixed \citet{Tinker2008} coefficients we have adopted for our halo mass function become unreliable at higher $z$. We do not attempt to correct for this offset because halos beyond $z\sim 3.5$ make a negligible contribution to the quantities of interest for us.

\subsection{Cosmic-ray injection and boundary conditions}
\label{ssec:CR_boundaries}

The second ingredient in our model, which sits on top of the halo evolution component, is a prescription for the injection of CRs into the CGM and their behaviour once they reach $\rvir$.

A well-defined virial shock is anticipated \citep{Ensslin2011} to define a trapping boundary for $\sim$GeV-scale CRs, so the CR outer boundary mirrors the three CGM scenarios of \autoref{sssec:cgm_model}. In \textsc{TrappingMax} and \textsc{TrappingMin} the virial shock sits at $R = 1$ and confines the CRs there (even in \textsc{TrappingMin}, where some CGM gas lies outside it), so we adopt a zero-flux boundary condition at $R=1$. In \textsc{FreeEscape} there is no virial shock and CRs escape freely into the IGM: the domain is the whole CGM, from $R = R_0$ to $R_\mathrm{clos}$, with a free-escape boundary at $R=R_\mathrm{clos}$. The precise mathematical form of each boundary condition is given in \autoref{ssec:transport}.

\subsubsection{CR injection and the inner boundary}
\label{sssec:CR_injection}

The inner boundary condition in our model describes the rate at which CR proton energy escapes from the ISM into the CGM. Here we adopt a fiducial relativistic CR proton energy release per core collapse supernova (CCSN) of $10^{50}$ erg, $\sim$10\% of the supernova's mechanical energy release \citep[e.g.,][and references therein]{Crocker2021a}.
If CCSN require progenitors of zero-age main sequence mass of at least 8 $\msun$ and we adopt a \citet{Chabrier2005} IMF, then the canonical CR-proton energy yield per unit stellar mass formed is
\begin{equation}
    \eta_c \equiv \frac{10^{50} \ {\rm erg}}{90 \ \msun} \simeq 1.1 \times 10^{48} \ {\rm erg} \ \msun^{-1} .
\label{eq:cr_injection_efficiency}
\end{equation}
The total, relativistic hadron CR power accelerated by ongoing star formation then scales with the star-formation rate as $\dot{E}_\mathrm{p} = \eta_c \mdotstar$.

The remaining quantity needed to finish specifying the boundary conditions is the calorimetry fraction $f_\mathrm{cal}$, defined as the  fraction of this power that is lost the dense galactic disc before it enters the CGM. In the framework of our model, $f_{\rm cal}$ is parametrised, at any particular epoch $z$, by $\{M_\star,\dot{M}_\star,r_{\rm d}\}$, the stellar mass, star formation rate, and disc radius (which affects the gas surface density) that we assign to a particular halo mass $\mh$. In reality $f_\mathrm{cal}$ is almost certainly a function of CR energy as well, but we will neglect this complication and adopt a single energy-averaged calorimetry fraction.

To construct our effective $f_\mathrm{cal}$, we start from the model implemented in the \congr{} code, which predicts the energy-dependent calorimetry fraction as a function of $\{M_\star,\dot{M}_\star,r_{\rm d}\}$, and which has been shown to faithfully reproduce the observed $\gamma$-ray spectra of local galaxies, the FIR-$\gamma$ and FIR-radio correlations, and the IGRB \citep{Roth2021, Roth2023, Roth2024}\footnote{The normalisation of the \citet{Roth2021} background component has been disputed \citep{Chen2025}. We defer discussion to Paper III.}. Our method for reducing this energy-dependent $f_\mathrm{cal}$ to a single effective number is as follows: for a given $\{M_\star,\dot{M}_\star,r_{\rm d}\}$, \congr{} predicts both the injected CR proton spectrum $d\dot N_{\rm inj}/dE_{\rm p}$ and the corresponding escaping CR proton power spectrum,
\begin{equation}
    E_{\rm p}\frac{d\dot N_{{\rm esc}}}{dE_{\rm p}}
    =
    E_{\rm p}\frac{d\dot N_{{\rm inj}}}{dE_{\rm p}}
    \left[
    1-f_{{\rm cal}}(E_{\rm p})
    \right].
\end{equation}
From these we define an effective energy-weighted escape fraction by introducing a spectral tilt parameter \(\delta\alpha\) and a renormalisation factor $q_\mathrm{cal}$, and then an effective energy-averaged calorimetry fraction
\begin{equation}
    f_\mathrm{cal}(\delta\alpha, q_\mathrm{cal}) = 
    q_\mathrm{cal} \frac{\int E_{\rm p} (d\dot{N}_\mathrm{inj}/dE_{\rm p}) f_\mathrm{cal}(E_{\rm p}) W(E_{\rm p},\delta\alpha)\, dE_{\rm p}}{\int E_{\rm p} (d\dot{N}_\mathrm{inj}/dE_{\rm p}) W(E_{\rm p},\delta\alpha) \, dE_{\rm p}},
    \label{eq:fcal_defn}
\end{equation}
where $W(E_{\rm p},\delta\alpha) \propto E_{\rm p}^{-\delta\alpha}$. A positive \(\delta\alpha\) gives greater weight to lower-energy CRs and tends to increase the effective disc calorimetry. The factor $q_\mathrm{cal}$ compensates for this by multiplying the overall energy-averaged calorimetry fraction by a factor that is slightly smaller than unity to keep the total disc contribution to the IGRB unchanged.

We calibrate the parameters $(\delta\alpha, q_\mathrm{cal})$ so that our prediction of the \textit{disc} contribution to the IGRB from 1 to 10 GeV agrees with the numerical predictions of \congr{}, rescaled to our own cosmic-ray yield and star-formation history; the full procedure (\aref{app:fcal_calibration}) yields $\delta\alpha = 0.15$ and $q_\mathrm{cal} = 0.77$, which we adopt throughout. With these choices, our simplified energy-independent $f_\mathrm{cal}$ reproduces the rescaled, fully energy-dependent \congr{} prediction to $\approx 1\%$ over the $1-10$ GeV energy range on which we focus here.

In principle once we have fixed our calibration of $(\delta\alpha, q_\mathrm{cal})$, we could use \autoref{eq:fcal_defn} directly to compute the inner boundary condition for each halo at each $z$ in our model since, for any given halo mass-redshift pair $(\mh,z)$, the model described in \autoref{ssec:halos} predicts $\{M_\star,\dot{M}_\star,r_{\rm d}\}$. In practice this approach is too computationally-expensive, as it would require us to run \congr{} for each case to generate the energy-dependent calorimetry fraction $f_\mathrm{cal}(E_{\rm p})$ that enters \autoref{eq:fcal_defn}. Instead, we adopt the approach of running \congr{} on the CANDELS galaxy sample \citep{Grogin2011, van_der_Wel2012} -- the same sampled used in the \citet{Roth2021} IGRB investigation -- to generate a 4D data set of $\{\log M_\star,\log \dot{M}_\star,\log r_{\rm d}, f_\mathrm{cal}\}$ points. We then use the \textsc{Mathematica} function \textsc{SmoothKernelDistribution} to generate a kernel density estimate of the data that we can use for rapid evaluation of $f_\mathrm{cal}$ at any $\{M_\star,\dot{M}_\star,r_{\rm d}\}$. The inner boundary condition  for any halo of mass $\mh$ at  $z$ is that the CR flux into the CGM is
\begin{equation}
    \dot{E}_{\rm p, inj} = (1 - f_\mathrm{cal}) \eta_c \mdotstar,
    \label{eq:dotEpinj}
\end{equation}
where $f_\mathrm{cal}$ is evaluated for the $\{M_\star,\dot{M}_\star,r_{\rm d}\}$ of that halo as predicted by the model in \autoref{ssec:halos}.

\subsection{CR transport model}
\label{ssec:transport}

The last model piece is a prescription for CR transport in the CGM. In what follows, subscript p denotes particle-level quantities (e.g., $E_{\rm p}$ for the energy of a single CR proton) and subscript c population-level quantities (e.g., $u_{\rm c}$ for the CR proton energy density).

\subsubsection{The transport equation}
\label{ssec:trans_eq}

We treat the CGM as spherically-symmetric, and assume that the transport of CRs can be described, at least approximately, as a diffusion process. This approximation is valid even if microphysically CRs stream rather than diffuse along the field lines, as we expect to be the case in the CGM where the waves emanate primarily from the galaxy -- or from the CRs themselves --and thus are highly anisotropic. This is because streaming plus field line random walk gives rise to approximately diffusive behaviour as long as magnetic fields are tangled on a scale smaller the size scale of the halo \citep[e.g.,][]{Sampson23a}, as predicted by simulations of CGM magnetic fields \citep[e.g.,][]{Pakmor2020}. 
However, we do retain a term to describe streaming losses.
Under this assumption, the evolution equation for the CR population is
\begin{equation}
    \pdv{u_{\rm c}}{t}
    =
    \frac{1}{r^2}\pdv{}{r}\left(r^2 D\pdv{u_{\rm c}}{r}\right)
    -
    \frac{u_{\rm c}}{t_{\rm pp}}
    -
    \frac{v_A}{3}\abs{\pdv{u_{\rm c}}{r}},
    \label{eq:uc_dimensional}
\end{equation}
where $r$ is the radial position within the CGM. The first term on the right-hand side describes diffusion with coefficient $D$, the second represents hadronic losses with timescale $t_\mathrm{pp}$, and the third is the streaming-loss rate, with $v_A$ as the Alfv\'en speed in the gas. As usual for spherical diffusion equations, it is convenient to make a change of variables by defining $v_c = r u_c$, reducing the equation to a one-dimensional linear diffusion equation,
\begin{equation}
    \pdv{v_{\rm c}}{t}
    =
    D\frac{\partial^2 v_{\rm c}}{\partial r^2}
    -
    \frac{v_{\rm c}}{t_{\rm pp}}
    -
    \frac{v_A}{3}\abs{\pdv{v_{\rm c}}{r}-\frac{v_{\rm c}}{r}}.
    \label{eq:vc_dimensional}
\end{equation}

We expect that $D$, $t_\mathrm{pp}$, and $v_A$ will all vary with redshift $z$, and that $t_\mathrm{pp}$ will vary with radius as the gas density varies. For $D$, we adopt a deliberately-agnostic approach as to the nature of CR transport by simply parametrising the diffusion coefficient as
\begin{equation}
    D(z)= \frac{1}{3} \Kdiff \rvir c_s,
    \label{eq:D_K}
\end{equation}
where $\Kdiff$ is a dimensionless parameter whose value we will vary and $c_s$ is the sound speed in the CGM, which we compute assuming that the CGM is at the virial temperature of the NFW halo. Thus $c_s = \sqrt{\gamma k_B T_\mathrm{vir}/(\mu m_{\rm p})}$, the adiabatic sound speed with $\gamma = 5/3$, where $\mu = 0.6$ is the mean molecular weight for fully ionised gas and the virial temperature follows from the virial theorem, $k_B T_\mathrm{vir}/(\mu m_{\rm p}) = |U_{\rm grav}|/(3\mh)$, with $U_{\rm grav}$ the gravitational self-energy of the NFW halo within $\rvir$ (see \aref{app:cgm_binding_energy}); for halo concentrations of interest $c_s=(5f_{\mathcal C}/9)^{1/2}(G\mh/\rvir)^{1/2}\simeq(0.85$--$0.95)\,(G\mh/\rvir)^{1/2}$, i.e.\ $c_s\simeq 104\,\kms$ for $\mh = 10^{12}\,\msun$ at $z=0$.
This parametrisation is convenient because it isolates the physically relevant multiplier of the natural halo-scale diffusivity $(1/3)\,\rvir c_s$ and is not tied to a particular microscopic model for CR transport.

The hadronic loss time, $t_\mathrm{pp}$, from our CGM density profiles is
\begin{equation}
    t_{\rm pp}=\frac{1}{n_e \sigma_{\rm pp}\kappa_{\rm pp}c}.
    \label{eq:tpp_dimensional}
\end{equation}
Here $n_e=\rho_{\rm g}/(\mu_e m_{\rm p})$ is the electron density, $\rho_{\rm g}$ is the gas density as a function of $r$ and $z$ given by \autoref{eq:cored_profile}, $\mu_e = 1.17$ is the mean mass per electron of a fully ionised plasma with hydrogen mass fraction $X\simeq 0.71$, $\sigma_\mathrm{pp}$ is the cross-section for hadronic interactions, and $\kappa_{\rm pp}$ is the inelasticity. For the latter two quantities we adopt the \citet{Nakamura2010} parametrisation of the \emph{total} {\it pp} cross-section evaluated at $E_p = 10$\,GeV, $\sigma_{\rm pp}\simeq 39.7$\,mb, with inelasticity $\kappa_{\rm pp}=1/2$ \citep[the same effective values adopted in][]{Crocker2021a}. Relative to a pure {\it pp}-on-hydrogen treatment using the inelastic cross-section \citep[$\simeq 30$\,mb at 10\,GeV;][]{Kafexhiu2014}, this combination raises the effective hadronic emissivity per unit gas mass by a factor ${\simeq}1.6$, placing it within ${\sim}10$ per cent of a full accounting of the nuclear enhancement from He and heavier species in both beam and target (${\simeq}1.8\times$ pure {\it pp}).

Finally, for $v_A$ we adopt $v_A/c_s = 0.1$, which corresponds to a plasma $\beta \equiv P_{\rm gas}/(B^2/8\pi) = 2c_s^2/(\gamma v_A^2) = 120$, consistent with numerical simulations of CGM magnetic fields in which the small-scale dynamo saturates at $\beta \sim 10^2$ \citep[e.g.,][]{Pakmor2017, Pakmor2020}. We write this ratio as $v_A/c_s = \Mturb/\MA$, where $\Mturb \equiv \delta v/c_s$ and $\MA \equiv \delta v/v_A$ are the sonic and Alfv\'enic Mach numbers of the CGM turbulence, with fiducial values $\Mturb = 0.3$ and $\MA = 3$; $v_A$ is taken to be constant with radius within any one halo. We explore variations in this choice, and show that the observational constraints in fact require $\beta$ to lie at approximately this value, in Paper III.

\subsubsection{Non-dimensionalisation and boundary conditions}
\label{sssec:non-dimensionalisation}

Our next step is to non-dimensionalise the transport equation as a prelude to solving it. The non-dimensionalisation requires some care because the boundaries of the problem are defined in terms of $\rvir$, which it itself evolving with time.

Our natural length and time scales for the non-dimensionalisation are the virial radius $\rvir$ and the diffusive timescale to reach the virial radius, $\rvir^2 / D$. Formally, we define
\begin{equation}
    r_\star \equiv \rvir
    \qquad
    t_\star \equiv \frac{r_\star^2}{D}
    =\frac{3}{\Kdiff}\,\frac{r_\star}{c_s},
    \label{eq:tstar_K}
\end{equation}
where we remind readers that $\rvir$ and $c_s$, and thus $r_\star$ and $t_\star$, are functions of both $\mhn$ and $z$. 
For our natural scale of energy density, which we require to non-dimensionalise $u_c$, we use the present-day binding energy of the CGM gas the CRs sample (i.e., excluding material outside $R=1$ in the \textsc{TrappingMin} case), $U_{\rm CGM,0}$, divided by the volume inside $r_{\star,0}$, both evaluated at the present day. We define a time-independent surface-energy scale and the corresponding redshift-dependent energy density
\begin{eqnarray}
    v_\star &\equiv& \frac{3U_{\rm CGM,0}}{4\pi r_{\star,0}^{2}},
   \nonumber \\
    u_\star(z) &=& \frac{v_\star}{r_\star(z)}, \nonumber \\
    U_{\rm CGM,0} &=& \frac{G\mhn M_{\rm CGM,0}}{r_{\star,0}}\,\mathcal{U},
    \label{eq:ustar_main}
\end{eqnarray}
where a zero in the subscript indicates that a particular quantity is evaluated at $z=0$, and $\mathcal{U}$ is a dimensionless function of order unity that depends on the choice of CGM boundary condition; we give an explicit form for this function in \aref{app:cgm_binding_energy}. 

With these choices, it is algebraically straightforward to rewrite \autoref{eq:vc_dimensional} as
\begin{equation}
    \pdv{V}{z}
    -
    \frac{R}{z_{r_\star}}\,\pdv{V}{R}
    -
    \frac{1}{\Zfac}\frac{\partial^2 V}{\partial R^2} 
    -
    \frac{\tau_s\,\tau_{\rm pp}}{\Zfac}
    \frac{V}{(R^2+R_0^2)^{\alpha/2}}
    -
    \frac{\tau_s}{\Zfac}\abs{\pdv{V}{R}-\frac{V}{R}}
    =0,
    \label{eq:V_maintext}
\end{equation}
where $R = r/r_\star$ is the dimensionless radius, $V = R u_c / u_\star$ is the dimensionless CR energy density times radius, the  factors
\begin{equation}
    \Zfac \equiv (1+z)t_\star H(z) \qquad 
    z_{r_\star} \equiv\left(\frac{1}{r_\star}\frac{dr_\star}{dz}\right)^{-1}
    \label{eq:Z_backbone}
\end{equation}
are dimensionless Jacobians arising from the transformation from $t$ to $z$ as our time variable, and the two factors
\begin{equation}
    \tau_s \equiv \frac{v_A}{\Kdiff c_s} = \frac{\Mturb}{\Kdiff\,\MA} \qquad
    \tau_{\rm pp} \equiv \frac{t_\star}{t_{\rm pp}} \frac{(R^2+R_0^2)^{\alpha/2}}{\tau_s}.
    \label{eq:taus_backbone}
\end{equation}
can be thought of as measurements of the effective optical depth of the CGM to streaming and hadronic losses, respectively. That is, $\tau_s$ parametrises the ratio of diffusive escape time to the time required for a population of CRs to dissipate their energy into heat by streaming, while $\tau_{\rm pp}$ measures the ratio of the diffusive escape to the hadronic loss times. Because the factor $(R^2 + R_0^2)^{\alpha/2}$ in $\tau_{\rm pp}$ cancels the corresponding dependence on radius in $n_e$ and thus in $t_{\rm pp}$ (c.f.~\autoref{eq:tpp_dimensional}), $\tau_{\rm pp}$ depends on $z$, but not on $R$.

We must solve \autoref{eq:V_maintext} subject to the inner and outer boundary conditions described in \autoref{ssec:CR_boundaries}. At the inner boundary we require that the CR flux $D (\partial u_c/\partial r)$ match that prescribed by CR injection, $\dot{E}_{\rm p,inj}/4\pi (R_0\rvir)^2$, where $\dot{E}_{\rm p,inj}$ is given by \autoref{eq:dotEpinj}. Converting this condition to our non-dimensional variables gives the Robin boundary condition
\begin{equation}
    \left.\pdv{V}{R}\right|_{R=R_0}
    =
    \left(
    \frac{\mathcal F_{\rm inj}}{R}
    -\frac{V}{R}
    -\frac{\Zfac}{z_{r_\star}}RV\right)_{R=R_0},
    \label{eq:BC_inner_robin}
\end{equation}
where
\begin{equation}
    \mathcal{F}_{\rm inj} = \frac{t_\star}{4\pi r_\star^3 u_\star} \dot{E}_{\rm p,inj}
\end{equation}
is the dimensionless CR energy injection rate.

At the outer boundary, for the \textsc{TrappingMin} and \textsc{TrappingMax} cases where we assume that a virial shock confines the CRs to the virial radius, we require that the CR flux $D (\partial u_c/\partial r)$ vanish at $r = \rvir$. For the \textsc{FreeEscape} case we assume that the CR energy density falls to zero at the edge of the CGM, with the flux across the boundary then determined by the resulting gradient in the CR energy density. Translating these conditions to our dimensionless variables we have
\begin{equation}
    \left.\pdv{V}{R}\right|_{R=1}
    =
    \left(
    \frac{V}{R}
    +\frac{\Zfac}{z_{r_\star}} RV\right)_{R=1}
    \label{eq:BC_outer_trap_robin}
\end{equation}
for \textsc{TrappingMin} and \textsc{TrappingMax}, and $V|_{R=R_\mathrm{clos}} = 0$ for \textsc{FreeEscape}.

\subsubsection{Numerical method}
\label{sssec:numerical_method}

We solve \autoref{eq:V_maintext} in \textsc{Mathematica} with a method-of-lines scheme and one-dimensional finite elements on the radial interval relevant to the chosen boundary mode. The radial mesh covers $[R_0,R_{\rm out}]$, with $R_{\rm out}=1$ for \textsc{TrappingMin} or \textsc{TrappingMax} boundaries and $R_{\rm out} = R_{\rm clos}$ for \textsc{FreeEscape} boundaries. To increase computational speed, we pre-calculate the various redshift-dependent parameters appearing in the equation -- such as $\tau_{\rm pp}$, $t_\star$, $\mathcal{F}_{\rm inj}$ -- on a common redshift grid, then interpolate while solving \autoref{eq:V_maintext}. All calculations begin at $z = 8$ with $V = 0$ throughout the domain, and continue to $z = 0$.\footnote{Since our inner boundary condition (\autoref{eq:BC_inner_robin}) prescribes $\partial V/\partial R$, there is a numerical difficulty in starting with an initial condition $V = 0$ everywhere but $\mathcal{F}_{\rm inj}$ non-zero, since our boundary condition would then require that $\partial V/\partial R$ change discontinuously at the boundary. We circumvent this difficulty by multiplying $\mathcal{F}_{\rm inj}$ by a ramp function that goes from 0 at $z = 8$ to 1 at $z = 7.8$, thereby providing a smooth startup with no discontinuities.}

From our numerical solutions we tabulate a series of useful diagnostic quantities at each redshift:
\begin{eqnarray}
    E_{\rm CGM} & = & 4\pi r_\star^3 u_\star\int RV\,dR,
    \label{eq:ECGM_post}
    \\
    \dot E_{\rm st}
    & = & 4\pi \frac{r_\star^3 u_\star}{t_\star}\tau_s
    \int R\abs{\pdv{V}{R}-\frac{V}{R}}\,dR
    \label{eq:Edotst_post}
    \\
    \dot E_{\rm pp}
    & = & 4\pi \frac{r_\star^3 u_\star}{t_\star} \tau_s\tau_{\rm pp}
    \int \frac{R\,V}{(R^2+R_0^2)^{\alpha/2}}\,dR
    \label{eq:Edotpp_post}
    \\
        \dot E_{\rm esc} & = & -4\pi\frac{r_\star^3 u_\star}{t_\star} \left(R\,\pdv{V}{R}\right)_{R=R_{\rm clos}}
\end{eqnarray}
These quantities are, respectively, the instantaneous CR energy in the CGM, and the instantaneous rates at which CR energy is lost to streaming, hadronic collisions, and escape from the CGM (with the final term non-zero only for \textsc{FreeEscape} boundaries). We also integrate the instantaneous loss rates over time to produce cumulative losses over the full duration of our calculation. Using these diagnostics, we find that our numerical method conserves total CR energy to better than 2\% in all cases.

For any given choice of the free parameters $\Kdiff$ and $\alpha$ describing the CR diffusion rate and CGM gas density profile, and a choice of boundary condition, we use the method above to generate solutions on a grid of present-day log halo mass $\log(\mhn/M_\odot)$ from 10 to 13 in steps of 0.25. This set of solutions then becomes the basis for our comparison to the empirical constraints, which we describe next.

\subsection{Predictions for the empirical constraints}
\label{ssec:model_constraints}

Once we have our grid of solutions over $\log\mhn$ for a given dimensionless diffusion coefficient $\Kdiff$, CGM density index $\alpha$, and boundary condition (\textsc{TrappingMin}, \textsc{TrappingMax}, or \textsc{FreeEscape}), we generate predictions for each of the empirical constraints as follows.

\subsubsection{The IGRB}
\label{sssec:model_prediction_igrb}
\label{subsec:igrb_constraints}
\label{subsubsec:igrb_external}

For the IGRB, our point of comparison is the integrated all-sky flux over the $1-10$ GeV band (\autoref{ssec:IGRB}). To predict this flux from our model, we sum three components: the contributions from external star-forming discs, $F_{\gamma,{\rm disk}}$, external CGMs, $F_{\gamma,{\rm CGM}}$, and the Milky Way's own CGM, $F_{\rm MW}$.

For both extragalactic contributions the all-sky flux is
\begin{equation}
    F_{\gamma}
    =
    \int dz
    \int dM_{\rm h,0}\,
    \frac{dn}{dM_{\rm h}}\frac{d\mh}{d\mhn}\,
    \frac{c}{H(z)(1+z)^2}
    f_{\rm bolo} L_{\gamma} 
    ,
    \label{eq:igrb_lightcone}
\end{equation}
where $dn/d\mh$ is the halo number density as a function of halo mass $\mh$ and redshift $z$ (\autoref{eq:halo_density_tinker}), $d\mh/d\mhn$ is determined for each halo's mass $\mh(\mhn,z)$ and redshift $z$ from our tracks of halo growth (\autoref{ssec:halos}), $L_\gamma$ is the gamma-ray luminosity of the disc or CGM, and $f_{\rm bolo}$ is a bolometric correction to account for the fraction of emission that falls into the $1-10$ GeV band. The factor $c/[H(z)(1+z)^2]$ combines the comoving volume element with the luminosity distance, $d_L=(1+z)\,d_C$.

We compute the disc $\gamma$-ray luminosity from our calibrated \congr{}~model as described in \aref{app:fcal_calibration}, while we compute the CGM luminosity for each combination of $\mh$ and $z$ as $L_{\gamma,\rm CGM} = \xi_\gamma \dot{E}_{\rm pp}$, where $\dot{E}_{\rm pp}$ is the {\it pp} loss rate at that $(\mh,z)$ obtained from our numerical integration of the model (\autoref{sssec:numerical_method}) and we adopt $\xi_\gamma = 1/3$, corresponding to the standard approximation that roughly one third of the hadronic loss power ultimately emerges as photons through $\pi^0$ decay. We compute the bolometric corrections from the normalised shape of the emission spectrum $s_\gamma(E_\gamma, \alpha_p)$ (with $\int s_\gamma(E_\gamma) \, dE_\gamma = 1$) as $f_{\rm bolo} = \int_{(1+z)E_0}^{(1+z)E_1} s_\gamma(E_\gamma)\, dE_\gamma$, where $E_0 = 1$ GeV and $E_1 = 10$ GeV. The spectral shape function $s_\gamma$ depends on the momentum index $\alpha_p$ of the protons responsible for driving the emission, defined so that the proton energy distribution is $dn/dE_{\rm p}\propto E_{\rm p}^{-\alpha_p}$. We compute $s_\gamma$ for a given $\alpha_p$ from the differential cross sections provided by \citet{Kamae2006} and \citet{Kelner2006}, as described in \aref{app:fcal_calibration}; we then numerically integrate the expression above to obtain $f_{\rm bolo}$ as a function of redshift. For the disc contribution to the IGRB, we adopt $\alpha_p = \alpha_{p,0} + \delta\alpha$ with $\alpha_{p,0} = 2.2$ as our adopted spectrum of the CRs at injection and $\delta\alpha = 0.15$ the spectral softening of the disc emission due to preferential escape of high-energy CRs, calibrated to the full \congr{}~energy-dependent escape model (\aref{app:fcal_calibration}). This energy-dependent escape, on the other hand, means that the spectrum of CRs escaping from galaxies into the CGM is, if anything, harder than the injected spectrum. We nonetheless adopt $\alpha_p = \alpha_{p,0} = 2.2$ when computing the bolometric correction for CGM emission. The bias this introduces is small and of known sign: a harder spectrum places a smaller fraction of its photon power in the $1-10$\,GeV band ($f_{\rm bolo}$ falls by $\approx15$--20 per cent between $\alpha_p = 2.2$ and 2.1), so if the escaping spectrum is harder than injection our choice slightly overestimates the CGM contribution to the IGRB.

\label{subsubsec:mw_cgm_igrb}
To compute the contribution from the Milky Way's own CGM, $F_{\rm MW}$, we first compute the volumetric, bolometric-corrected $\gamma$-ray emissivity of the CGM as a function of dimensionless radius $R$ as (c.f.~\autoref{eq:Edotpp_post})
\begin{equation}
    q_\gamma(R) = f_{\rm bolo} \xi_\gamma \frac{u_\star}{t_\star} \tau_s \tau_{\rm pp} \frac{V(R)}{R(R^2 + R_0^2)^{\alpha/2}},
    \label{eq:q_gamma}
\end{equation}
where $u_\star$ and $t_\star$ are the dimensional scaling energy and timescale (\autoref{sssec:non-dimensionalisation}), $R_0 = 0.018$ is our fiducial dimensionless radius for the inner edge of the CGM, $V(R)$ is our numerical solution for the dimensionless CR energy density, and $\alpha$ is our adopted CGM density powerlaw index. In these calculations we adopt $\mhn = 10^{12}$ M$_\odot$ for the present-day mass of the Milky Way halo \citep{Wang2020}, and we calculate all quantities at $z=0$.

To convert this emissivity to a mean all-sky flux, we put  the Sun at a dimensionless distance $R_\odot = r_\odot/r_\star$ from the Galactic centre (with $r_\odot = 8.2$ kpc; \citealt{GRAVITY-Collaboration19a}). Under this assumption, the intensity observed at Earth along a line of sight at Galactic latitude $b$ and longitude $\ell$ is $I_\gamma = \frac{r_\star}{4\pi} \int q_\gamma(R(S)) \, dS$, where $S$ is the distance along the line of sight,
$R(S) = (R_\odot^2 + S^2 - 2 R_\odot S \mu )^{1/2}$,
and $\mu = \cos b \cos\ell$ is the cosine of the angle between the line of sight and the Galactic centre; we implicitly take $q_\gamma(R) = 0$ for $R < R_0$ and $R> R_{\rm out}$, where $R_{\rm out} = 1$ for \textsc{TrappingMin} or \textsc{TrappingMax} boundaries, and $R_{\rm clos}$ for \textsc{FreeEscape} boundaries. Qualitatively, because of the Sun's displacement from the Galactic Centre, $I_\gamma$ is a monotonically decreasing function of $\mu$.

It is tempting at this point to define $F_{\rm MW}$ simply by integrating $I_{\gamma}$ over the full sky as seen from the Sun, but doing so would ignore important biases introduced by the way that the IGRB is measured from raw \textit{Fermi}/LAT data and how these procedures interact with the dependence of $I_\gamma$ on $\mu$. First, estimates of the IGRB mask the region with $|b| < b_{\rm mask} = 20^\circ$ to avoid the bright contribution from the Galactic Plane, thereby also excluding the region where $I_\gamma$ is largest. Second, even after excluding this region, estimates of the IGRB use templates that decompose the emission into an isotropic component plus a Galactic foreground, and since $I_\gamma$ is largest towards the inner Galaxy some of the flux from the Milky Way CGM may be absorbed into the foreground templates. For this reason we compute two estimates for $F_{\rm MW}$ that act as upper and lower bounds:
\begin{align}
    F_{\rm MW}^{\rm mono} & = \frac{4\pi}{\Omega_{\rm unmask}} \int_{|b|>b_{\rm mask}} I_\gamma \, d\Omega \\
    F_{\rm MW}^{\rm min} & = 4\pi \min_{|b|>b_{\rm mask}} I_\gamma,
\end{align}
where $\Omega_{\rm unmask} = 4\pi (1 - \sin b_{\rm mask})$ is the solid angle of the unmasked portion of the sky. The estimate given by $F_{\rm MW}^{\rm mono}$ corresponds to assuming that none of the large-scale gradient is absorbed into the foreground template and is thus an upper limit, while the lower limit $F_{\rm MW}^{\rm min}$ corresponds to assuming that only the minimum intensity present in any direction is retained, with the rest absorbed into the template.
The two limits nearly coincide in \textsc{FreeEscape} and differ by a factor $\simeq3.4$ in the trapping modes.

\label{subsubsec:igrb_benchmark}
To be deliberately conservative, we use $F_{\rm MW}^{\rm min}$ both when determining which parts of parameter space the IGRB excludes and when quoting the predicted circumgalactic share of the IGRB (\autoref{ssec:igrb_bbudget}).
Formally, we therefore consider a particular combination of parameters $(\Kdiff, \alpha)$ and CGM boundary condition to be ruled out if
\begin{equation}
    F_{\gamma,{\rm disk}} + F_{\gamma,{\rm CGM}} + F_{\rm MW}^{\rm min} > F_{\rm allow}^{\rm IGRB},
\end{equation}
where $F_{\rm allow}^{\rm IGRB}$ is our conservative upper limit given by \autoref{eq:igrb_1_10_allow}.

\subsubsection{CGM X-ray luminosities}
\label{sssec:model_prediction_xray}
\label{subsec:cgm_consistency}
\label{subsubsec:streaming_cooling}

CR streaming losses are dissipated as heat in the CGM, so in energy equilibrium $\dot E_{\rm st}$ (\autoref{eq:Edotst_post}) should not exceed the CGM bolometric cooling luminosity. Before comparing this to the observed X-ray luminosity, however, we must make a bolometric correction, since the observed luminosity only includes photons with energies in the range \(0.5\)–\(2\) keV. Following \citet{Mineo2012}, we adopt a factor $f_{\rm bolo} = 1/3$ for this purpose, and we therefore take $L_{\rm X,CGM} = f_{\rm bolo} \dot E_{\rm st}$ as our model prediction for the CGM X-ray luminosity.

We remind readers that our estimate of $\dot{E}_{\rm st}$ relies on the assumption that CRs in the CGM stream through the gas, and thus suffer losses at a rate \(\propto v_A\,\abs{\partial P_{\rm c}/\partial r}\). This in turn requires the presence of a net flux of Alfv\'en waves co-streaming with the CRs; under balanced extrinsic
turbulence, or purely magnetostatic scattering, the thermostat does not apply and this
constraint lifts. 

\subsubsection{$\gamma$-ray emission from the outer Milky Way}
\label{sssec:model_prediction_mw_rings}
\label{subsubsec:MW_ring}

The ring emissivities of \autoref{ssec:mw_profile} probe the local CR energy density directly, rather than a projected surface brightness, so we compare them with the model's CGM CR energy density $u_{\rm c} = u_\star V/R$ (\autoref{eq:ustar_main}) at $r\simeq8$\,kpc, for $\mhn = 10^{12}\,\msun$ at $z=0$. We apportion $u_{\rm c}$ between the two bands using the fractions of the energy density of the adopted $\alpha_p = 2.2$ CGM proton spectrum that fall in each, counted, like our normalisation of $\eta_c$, above the pion-production threshold ($\simeq0.26$ and $\simeq0.65$). While we compare the model-predicted CR energy densities to the the maximum allowed value in both the $1-10$ GeV and $>10$ GeV bands, in practice the harder band is the operative constraint for the spectrally hard CGM component. 

\subsubsection{CGM pressure balance}
\label{sssec:model_prediction_pressure}
\label{subsubsec:hydrostatic_consistency}

To evaluate the hydrostatic ratio $\eta_P$ (\autoref{eq:etaP_def}) we reconstruct the CR pressure from the transport solution, $P_{\rm c} = u_{\rm c}/3$ with $u_{\rm c}(r,z) = u_\star(z)\,V(z,R)/R$ (\autoref{eq:ustar_main}), and compare its gradient with the gradient $\rho_{\rm g}\,g$, $g = GM(<r)/r^2$, required to support the adopted gas profile (\autoref{eq:cored_profile}) in the halo potential. This is a stricter test than a global energetic argument because it asks whether the modelled CR population would over-support the gas locally at any radius. When $\eta_P\ll1$ the CR pressure gradient is dynamically subdominant and a hydrostatic gas configuration is self-consistent; when $\eta_P>1$ over an extended radial range 
the model has formally broken down.

\subsubsection{Virial shock confinement}
\label{subsubsec:confinement_prediction}

For the \textsc{TrappingMin} and \textsc{TrappingMax} boundary conditions, we also require that the CR pressure at  $r_{\rm vir}$ not exceed the ram pressure of the gas accreting onto the halo, $P_{\rm c}(r_{\rm vir})\le P_{\rm ram}$ (\autoref{ssec:pressure_support}), and we rule out any model that does not obey this constraint. We evaluate $P_{\rm ram}$ for smooth, spherically symmetric infall at $r_{\rm vir}$. Gas arriving with the cosmic baryon fraction at the rate $f_{\rm b}\dot M_{\rm h}$ and speed $v_{\rm vir}$ has density $\rho_{\rm in}=f_{\rm b}\dot M_{\rm h}/(4\pi r_{\rm vir}^2 v_{\rm vir})$, and so exerts ram pressure
\begin{equation}
    P_{\rm ram}=\rho_{\rm in}v_{\rm vir}^{2}
    =\frac{f_{\rm b}\,\dot M_{\rm h}\,v_{\rm vir}}{4\pi\,r_{\rm vir}^{2}}.
    \label{eq:Pram}
\end{equation}
Here we take $\dot M_{\rm h}$ from \autoref{eq:mh12dot_backbone}, and $v_{\rm vir}\equiv(G\mh/r_{\rm vir})^{1/2}$ is the circular velocity at the virial radius ($c_s/v_{\rm vir}\simeq0.9$; \autoref{ssec:trans_eq}). For an $L_\star$ halo at $z\simeq0$ the model's own accretion rate ($\sim6\,\msun\,{\rm yr}^{-1}$ of baryonic mass, consistent with \citealt{Fakhouri2010}) gives $P_{\rm ram}\simeq3.7\times10^{-16}\,{\rm dyn\,cm^{-2}}$. 

\section{Results}
\label{sec:results}

Having explained how we generate our library of models and evaluate the observational constraints for them, we now present our results in order of increasing constraint complexity. We begin in \autoref{subsec:results_indiv_halos} by orienting the reader by walking through  some representative outputs of our model for individual halos. Then in \autoref{subsec:parameter_space_scans} we describe our survey of parameter space, sequentially adding each observational constraint and identifying what each rules out. 

\subsection{Representative cosmic ray profiles}
\label{subsec:results_indiv_halos}

\begin{figure*}
\centering
\begin{minipage}[t]{0.76\textwidth}\vspace{0pt}
\includegraphics[width=\linewidth]{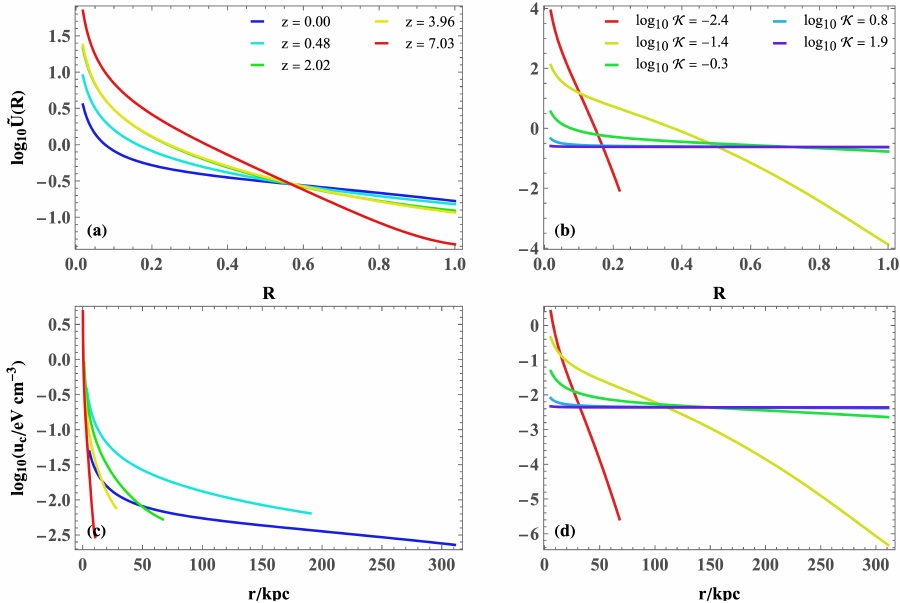}
\end{minipage}\hfill
\begin{minipage}[t]{0.225\textwidth}\vspace{0pt}\raggedright\scriptsize
\caption{The unit-normalised radial profile shape $\tilde U(z,R)$ of the CR energy density, as a function of $R = r/r_\star(z)$ (top row, panels a and b), and the corresponding dimensional energy density $u_{\rm c}$ as a function of $r$ (bottom row, panels c and d), for the boundary mode \textsc{TrappingMax}, for the example case of a CGM gas profile index $\alpha = 2.9$ and a $M_{\rm h,0} = 10^{12} \msun$ halo.
Each curve in the top row is separately normalised to unit energy content (\autoref{eq:Utilde_def}), so panels (a) and (b) carry profile \emph{shape} only; panels (c) and (d) show the corresponding amplitudes.
Left column (a, c): the profile at fixed, dimensionless transport coefficient $\log_{10}(\mathcal{K}) \simeq -0.27$ and for redshifts $z \in \{0.01, 0.5, 2, 4, 7\}$, indicated by the curve colours from blue to red (legend in panel a);
right column (b, d): the profile at fixed $z = 0$ and for various transport coefficients $\log_{10}(\mathcal{K})$, as indicated by the curve colours (legend in panel b).
}
\label{fig:plotrad1}
\end{minipage}
\end{figure*}

\begin{figure*}
\centering
\begin{minipage}[t]{0.76\textwidth}\vspace{0pt}
\includegraphics[width=\linewidth]{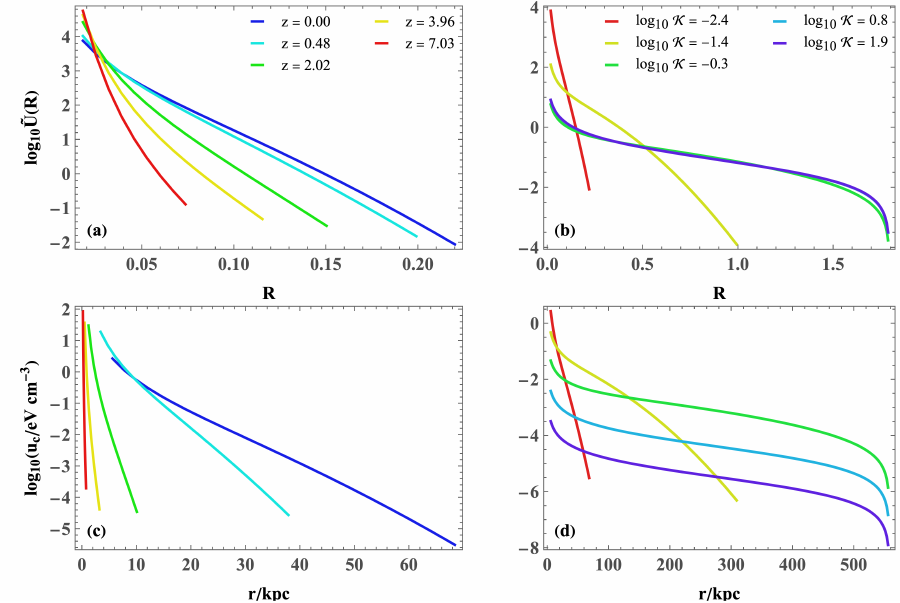}
\end{minipage}\hfill
\begin{minipage}[t]{0.225\textwidth}\vspace{0pt}\raggedright\scriptsize
\caption{Same as \autoref{fig:plotrad1}, with identical curve colour scheme and legends, but now for boundary mode
\textsc{FreeEscape} and for the example parameter
$\log_{10}(\mathcal{K})$ = -2.43 in the left column.
Note that, for \textsc{FreeEscape}, the profile can extend beyond $R = 1$.
}
\label{fig:plotrad2}
\end{minipage}
\end{figure*}

We show in \autoref{fig:plotrad1} and \autoref{fig:plotrad2} plots of the radial profile of the cosmic ray energy density in dimensional and non-dimensional terms for various representative parameter choices as returned by our \textsc{Mathematica} code. For the non-dimensional cases shown in the top rows of \autoref{fig:plotrad1} and \autoref{fig:plotrad2}, rather than showing $U = u_c/u_\star$, we instead show
\begin{equation}
    \tilde U(z,R) \;\equiv\;
    \frac{U(z,R)}{4\pi\!\int_{R_0}^{R_{\rm out}}\! R'^2\,U(z,R')\,{\rm d}R'}.
    \label{eq:Utilde_def}
\end{equation}
This removes normalisation differences between redshifts, making changes in the radial distribution easier to see.

Going from high (shown in red) to low (shown in blue) redshift, the left column figures show how, at fixed diffusion coefficient, the radial CR energy density evolves, typically from relatively steep to progressively flat.
The right column figures compare  profiles at $z \simeq 0.0$ for different diffusion coefficients.
As might be anticipated, profiles evolving under comparatively smaller diffusion coefficients remain more compact and steep with a larger energy density at $R_0$.
For the largest values of $\mathcal{K}$, the profiles are very broad and flat in the case where there is a trapping boundary, and the energy density becomes independent of $\mathcal{K}$. By contrast in the case of a \textsc{FreeEscape} boundary, some radial gradient in the profile is maintained even for large $\mathcal{K}$, and the CR amplitude drops as $\mathcal{K}$ rises.
This latter boundary condition also allows, in general, the profile to extend beyond $R=1$ to $R_{\rm clos}$ where it goes to zero by construction.

We can understand these results intuitively. Under \textsc{Trapping} nothing escapes, and the only sinks are the losses within the halo, hadronic collisions and streaming. As $\Kdiff$ rises, both become small: as the diffusion time becomes short, the interior of the halo homogenises, $u_{\rm c}\to{\rm const}$. This in turn suppresses both streaming and hadronic losses -- the former because there can be no streaming losses when there is no pressure gradient to stream down (loss rate 
$\propto\abs{\partial P_{\rm c}/\partial r}$, and the former because CRs rapidly escape from the denser CGM gas near the halo into the low-density outer halo. Consequently at $\Kdiff\gtrsim5$ the two together remove only a few
per cent of the injected energy, so nearly all of the CR energy injected over all cosmic time remains in the CGM at $z\simeq0$. Further raising $\Kdiff$
changes neither the normalisation nor the arrangement.
By contrast for \textsc{FreeEscape}
the shape becomes independent of $\mathcal{K}$ for large $\mathcal{K}$, but the amplitude does not. In this limit the CR profile is set by the balance between injection and escape, and that balance goes on shifting with $\Kdiff$. 

\subsection{Parameter-space scans}
\label{subsec:parameter_space_scans}

Having built up some intuition for how the solutions behave, we  proceed to generate solutions on a tensor product grid of the three possible boundary conditions (\textsc{FreeEscape}, \textsc{TrappingMin}, and \textsc{TrappingMax}), CGM density slope $\alpha = 1.1 - 2.9$ in steps of 0.1, and $\mathcal{K}$ sampled uniformly in logarithm over 33 values from $\approx 10^{-2.5}$ to $\approx 10^2$, corresponding to a physical range $D = 10^{28.0} - 10^{32.5}$ cm$^2$ s$^{-1}$ for a $10^{12} \msun$ halo at $z = 0.0$. We therefore have a total of 1881 parameter space samples, each of which is evaluated at 13 halo masses from $\log(\mhn/M_\odot) = 10 - 13$ as noted in \autoref{sssec:numerical_method}.

\subsubsection{The \textsc{TrappingMax} case}
\label{subsubsec:constrain_tmax}

\begin{figure*}
\centering
\begin{minipage}[t]{0.62\textwidth}\vspace{0pt}
\includegraphics[width=\linewidth]{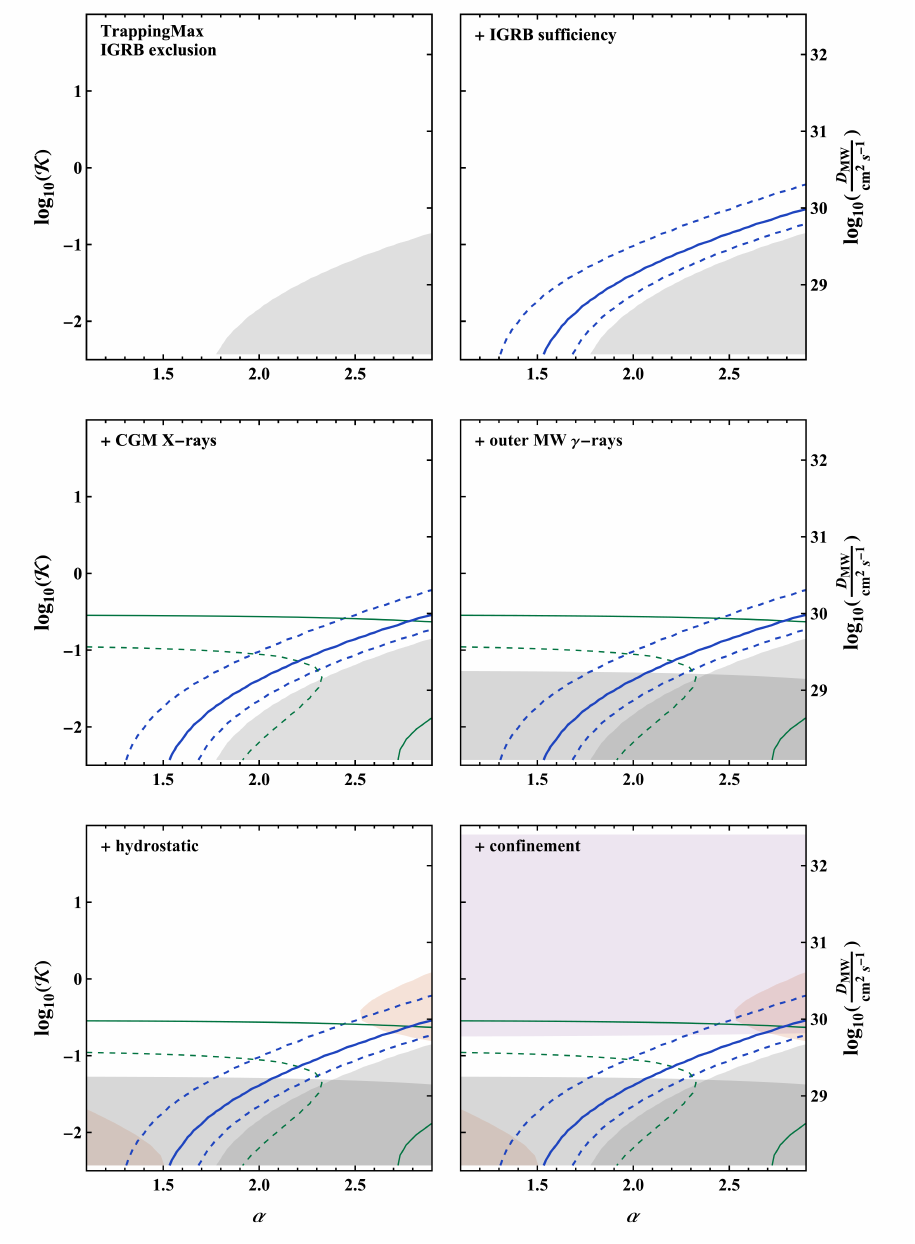}
\end{minipage}\hfill
\begin{minipage}[t]{0.365\textwidth}\vspace{0pt}\raggedright\small
\caption{Observational constraints applied cumulatively in the $(\alpha,\log_{10}\Kdiff)$ plane for the
\textsc{TrappingMax} boundary condition; the right vertical axis shows the corresponding physical diffusion coefficient for a Milky Way-mass halo at $z=0$, $\mhn = 10^{12}$ M$_\odot$. In this figure excluded regions are shown by shading, and sufficiency conditions by solid lines for central values with $\pm 1\sigma$ regions indicated by dashed lines in the corresponding colour.
Starting from the top left and proceeding across and down, we progressively add the 1--10\,GeV IGRB exclusion (light grey); the IGRB sufficiency condition (blue lines solid and dashed); the condition that the streaming losses fully explain CGM X-ray emission (green lines solid and dashed lines; the $-1\sigma$ sufficiency line and $+2\sigma$ exclusion region both lie outside the plot window); the region excluded by the observed CR energy density in the outer Milky Way (grey, across the bottom); the excluded region where $\eta_P > 1$ and thus the CGM cannot be hydrostatic due to CR pressure (orange); and the condition that the CR pressure at the virial radius not exceed the ram pressure of cosmological infall (purple, above $\Kdiff\approx0.2$).
The various exclusions clearly require $\Kdiff\simeq0.05$--$0.2$; over most of this range CRs are sufficient to both explain the X-ray luminosity of the CGM (area between solid green and dashed green lines) and, for $\alpha \gtrsim 1.8$, the IGRB (area between the two blue dashed lines).
}
\label{fig:constraintPlateB}
\end{minipage}
\end{figure*}

We first start with results for the \textsc{TrappingMax} boundary condition to gain some intuition before moving on to the other boundary conditions. \autoref{fig:constraintPlateB} shows the results of applying the observational constraints to the parameter space for this boundary mode one at a time, so that their separate effects can be followed.

Starting at the top left panel, the requirement that the model not exceed the 1--10\,GeV IGRB (\autoref{subsec:igrb_constraints}) excludes a region in the lower right-hand corner, where transport is slow and the gas profile is steep and therefore highly centrally concentrated. Slow transport keeps the CRs in the dense inner halo, where the target gas is concentrated, so the predicted hadronic emission from external CGMs and from the Milky Way's own CGM is greatest there. The top right panel then adds the region where the model satisfies the IGRB sufficiency condition outlined in \autoref{ssec:IGRB}; everywhere between the dashed blue lines  (with the solid line marking the central value of the measured IGRB), the $\gamma$-ray emission produced by a combination of extragalactic discs, extragalactic CGMs, and the Milky Way's own CGM explains the observed IGRB to within $\pm 1\sigma$.

We next consider the constraint from CGM X-ray emission (\autoref{subsubsec:streaming_cooling}), which proves to be most constraining for halos with mass $\mhn \approx 10^{12}$ M$_\odot$. We find that, even for these halos, the heating of the CGM by streaming never yields X-ray luminosities that exceed the observed one by $>2\sigma$ over our parameter space, and thus none of space is excluded by X-ray data. However, we \textit{do} find a region where CGM heating is sufficient to explain the observed X-ray emission within $\pm 1\sigma$, which we mark by the green lines in the centre left panel.
(A second heating--cooling balance contour lies at the far lower right of parameter space, but is wholly within the region already excluded by the IGRB.)

The fact that the model reproduces the observed CGM X-ray luminosity within $2\sigma$ everywhere over our broad parameter space is a reflection of the fact that the streaming loss rate depends only weakly on $\mathcal{K}$, largely due to a fortuitous cancellation; when $\mathcal{K}$ is large the streaming loss rate is suppressed because the gradient flattens, as discussed in \autoref{subsec:results_indiv_halos}, and so the streaming loss rate increases with $\mathcal{K}$, but only up to a point, because when $\mathcal{K}$ is small CRs are confined close to the galaxy where the gas is dense, and hadronic losses begin to dominate, setting an upper limit on how much of the CR power is ultimately  channelled into heating and thence X-rays.

Adding the constraints on the Milky Way circumgalactic CR energy density from the \textit{Fermi} ring
analyses of the outer Galaxy (\autoref{sssec:model_prediction_mw_rings}) removes a horizontal swathe across the bottom of
the plotted parameter space (grey; centre-right panel). The mechanism is that if CRs are transported too slowly then they remain concentrated in the inner halo, and the CGM contribution to the CR energy density at $r\simeq8$\,kpc then exceeds the measured value --- which, the rings being dominated by disc CRs, bounds that contribution from above (\autoref{subsubsec:MW_ring}). Because these limits bound the CR energy density directly rather than a $\gamma$-ray flux, they depend only weakly --- through the hadronic losses --- on the adopted CGM gas density profile. And because they measure one halo --- our own --- rather than a sum over the halo population,
the same limit is imposed nearly unchanged in every boundary mode.

We next add the constraint $\eta_P < 1$, so that the CGM can be approximately hydrostatic -- as expected for $L_\star$ and larger halos \citep[e.g.][]{Faerman2017,Faerman2020,Stern2019,Singh2024}. The
region where $\eta_P > 1$ anywhere within the CGM and hydrostatic balance is impossible indicated by the orange shading in the bottom left panel.
We find two distinct regions that are forbidden by the hydrostatic balance condition: a corner at $\alpha\lesssim1.5$ and low $\Kdiff$, and a wedge at $\alpha\gtrsim2.5$ and intermediate $\Kdiff$.
These regions have distinct origins: the low-$\alpha$, low-$\mathcal{K}$ corner becomes unstable at $\sim 0.05\rvir$  because this region of parameter space combines slow CR transport, which maximizes the CR pressure gradient in the inner halo, with a flat gas density profile, which minimizes gravitational force in the same region. By contrast, the high-$\alpha$, intermediate-$\mathcal{K}$ part of parameter space violates the hydrostatic constraint much closer to $\rvir$. This region appears because at high $\alpha$ baryons are concentrated in the inner CGM, lowering the gravitational force at larger radii, while for intermediate $\mathcal{K}$ CRs escape quickly from the dense inner region without suffering too many losses, but still maintain a pressure gradient rather fully flattening as happens at still higher $\mathcal{K}$.

Finally, because we are showing the constraints for a trapping boundary condition, we have the requirement that the ram pressure of the virial shock be sufficient to confine the CRs (\autoref{subsubsec:confinement_prediction}). This requirement fails to be met throughout the purple region (lower right panel), which begins at $\Kdiff\approx0.2$. Within this region the CR pressure at $r_{\rm vir}$ exceeds $P_{\rm ram}$ by a factor $3$--$6$, and the reflecting boundary the solution assumes could not be maintained. Within the surviving region, by contrast, $P_{\rm c}(r_{\rm vir})\le P_{\rm ram}$ by construction; confinement sets the region's highest point for $2.3\lesssim\alpha\lesssim2.7$, while at lower $\alpha$ the IGRB sufficiency condition lies below it.

\subsubsection{Constraints for all boundary modes}

\begin{figure}
    \centering
    \includegraphics[width=0.95\columnwidth]{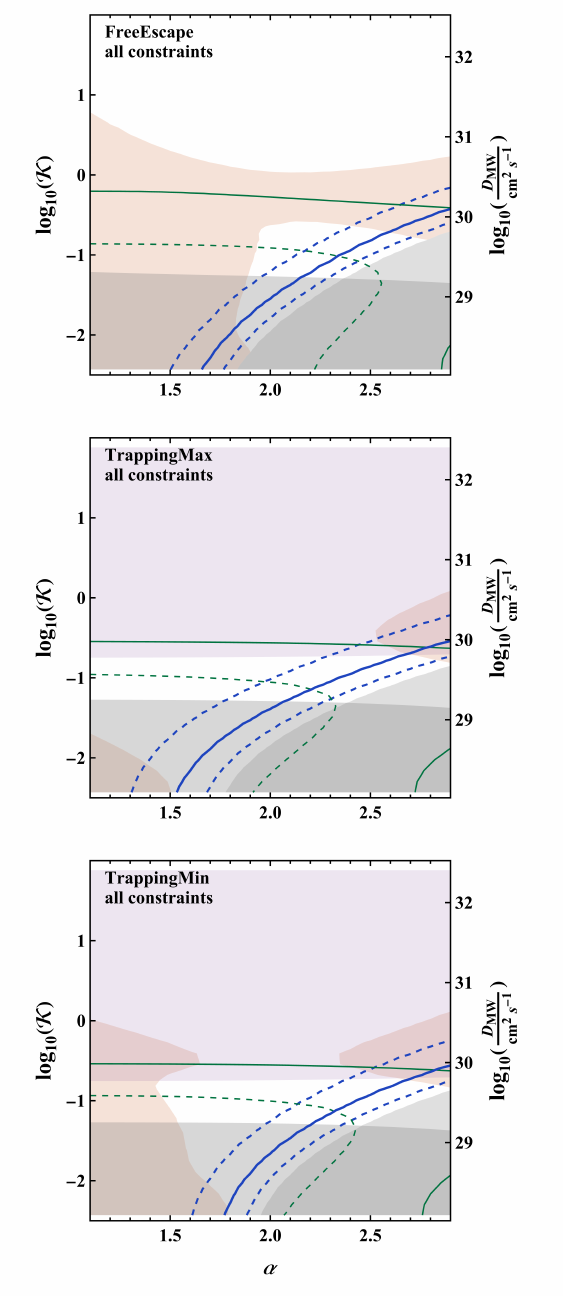}
    \caption{Same as the final panel of \autoref{fig:constraintPlateB}, but now showing constraints for each of the three boundary conditions, as indicated. The shaded regions and curves have the same meanings as in \autoref{fig:constraintPlateB}.}
    \label{fig:constraintPlateC}
\end{figure}

We now repeat the exercise above for every boundary mode and show the result in
(\autoref{fig:constraintPlateC}). Several features stand out.

First, there is a firm, mode-independent lower limit $\Kdiff\gtrsim0.05$ set by the \textit{Fermi} ring limits at low $\alpha$ and by IGRB overproduction at high $\alpha$. The former is only weakly dependent on the CGM gas profile, the latter on the halo population, and
neither is particularly sensitive to the boundary condition. The fundamental reason for this insensitivity is that these constraints mainly apply in the inner CGM -- the region from which the \textit{Fermi} ring limits come and which dominates the IGRB contribution owing to its high gas density -- where the solution is insensitive to boundary conditions.

The situation with respect to upper limits is more complicated and boundary-dependent. Most obviously, the requirement that the CR pressure not exceed infall ram pressure in the two \textsc{Trapping} cases, which eliminates most of the parameter space at high $\mathcal{K}$, does not apply if we assume that CRs prevent the formation of a virial shock entirely and can thus can freely escape into the IGM. More subtly, we see that the two regions that are forbidden on the grounds of hydrostatic balance in the \textsc{TrappingMax} case become larger in \textsc{TrappingMin}, and still larger and overlapping in \textsc{FreeEscape}. The instability regions expand in these cases simply because for these boundary conditions there are fewer baryons inside the virial radius. This both weakens the stabilising gravitational force and reduces CR hadronic losses. The effect is largest for  \textsc{FreeEsacpe}, which retains the fewest baryons within $\rvir$, dropping to only 5\% of the CGM mass residing within $\rvir$ for $\alpha = 1.1$.

We therefore see that there are two classes of solution permitted by the observations. One exits only for the \textsc{FreeEscape} boundary condition, and corresponds to very large $\mathcal{K}$. The physical nature of this solution is that CRs rapidly escape into the IGM, driving negligible $\gamma$-ray or X-ray emission and providing negligible pressure. For this solution CRs are essentially irrelevant to the CGM.

The other class of solution is far more interesting. This region, characterised by $\mathcal{K} \simeq 0.05 - 0.2$ and $\alpha\simeq 2 - 2.9$ and allowed by all three boundary modes, is characterised by a suggestive confluence. Not only is it \textit{allowed} by all the observational constraints, in this part of parameter space CRs in the CGM naturally explain both the heating of the CGM in $L_\star$ galaxies, as constrained by the \textit{eRosita} observations, and the IGRB, as constrained by \textit{Fermi} observations. We emphasise that there is no \textit{a priori} reason that these conditions should have coincided: the Milky Way ring analysis that rules out the low $\mathcal{K}$ part of parameter space and the hydrostatic balance and confinement conditions that rule out the high $\mathcal{K}$ parts are completely unrelated to the IGRB or CGM X-ray luminosity. The fact that all these unrelated constraints and sufficiency conditions overlap in the same narrow swathe of parameter space is therefore very interesting.

\section{Discussion}
\label{sec:discussion}

Having identified the CR transport regimes and CGM physical conditions that a ruled out by various observational constraints, we next work through the implications of these restrictions for quantities that are not directly observable. We examine the total CR energy budget in \autoref{ssec:CR_budget}, the contributions of different sources to the IGRB in \autoref{ssec:igrb_bbudget}, the role of CRs in providing heating and pressure support to the CGM in \autoref{ssec:cgm_budget}, and the nature of CR transport in the CGM in \autoref{ssec:cgm_transport}.

Here we will find it helpful to subdivide the remaining, allowed parameter space shown in \autoref{fig:constraintPlateC} into distinct regions. For the \textsc{TrappingMax} and \textsc{TrappingMin} boundary conditions, we distinguish between the entire allowed parameter space (the full unshaded region in the lower two panels of \autoref{fig:constraintPlateC}) and the parts of that parameter space that meet the IGRB sufficiency condition (the part of the unshaded region that falls between the two blue dashed lines); we refer to these regions as \textsc{TrappingMax[all]} and \textsc{TrappingMax[IGRB]}, and similarly for \textsc{TrappingMin}. Since there is near-complete overlap between the parts of parameter space that meet the IGRB sufficiency condition and those that meet the X-ray sufficiency condition (region above the dashed green line), we do not bother to distinguish meeting both conditions as a separate case. For \textsc{FreeEscape}, in addition to these two regions (\textsc{FreeEscape[all]} and \textsc{FreeEscape[IGRB]}), corresponding to the lower unshaded region in the top panel of \autoref{fig:constraintPlateC}, we also have the upper unshaded region of large $\mathcal{K}$, to which we refer as \textsc{FreeEscape[high]}; we discuss the physical meaning and plausibility of this particular range of parameter space further in \autoref{ssec:high-K}. We summarise the range in $(\alpha,\mathcal{K})$ to which each of these regions approximately corresponds in \autoref{tab:cases}.

\begin{table*}
    \begin{tabular}{ccccccccc}
        \hline\hline
        Quantity &
        \multicolumn{7}{c}{Parameter space regions}
        \\
        &
        \textsc{TMax[all]} & 
        \textsc{TMax[IGRB]} & 
        \textsc{TMin[all]} & 
        \textsc{TMin[IGRB]} &
        \textsc{FE[all]} & 
        \textsc{FE[IGRB]} &
        \textsc{FE[high]}
        \\ \hline
        \multicolumn{8}{c}{Defining parameters}
        \\ \hline
        $\alpha$ & (1.1, 2.9) & (1.8, 2.9) & (1.5, 2.9) & (2.0, 2.9) & (1.1, 2.9) & (2.0, 2.7) & (1.1, 2.9) \\
        $\log\mathcal{K}$ & $(-1.31,-0.72)$ & $(-1.31,-0.72)$ & $(-1.31,-0.72)$ & $(-1.31,-0.72)$ & $(-1.29,-0.59)$ & $(-1.29, -0.65)$ & $(0.03,1.88)$
        \\ \hline
        \multicolumn{8}{c}{Percent of CR energy injected into an $L_\star$ CGM over all time that by $z=$ has gone into$\ldots$}
        \\ \hline
        Hadronic loss & $(2.1,24)$ & $(9.4, 24)$ & $(13, 24)$ & $(8.9, 24)$ & $(2.3,16)$ & $(5.6, 16)$ & $<1$ \\
        Streaming loss & $(34,63)$ & $(33,58)$ & $(35,62)$ & $(34,58)$ & $(37,61)$ & $(38,60)$ & $(0.02,18)$ \\
        Escape to IGM & - & - & - & - & $<1$ & $<1$ & $(39,100)$ \\
        Still in CGM & $(27,57)$ & $(27,54)$ & $(28,58)$ & $(27,54)$ & $(30,59)$ & $(31,55)$ & $(0.2,14)$
        \\ \hline
        \multicolumn{8}{c}{Percent of the observed IGRB contributed by$\ldots$}
        \\ \hline
        Ex-gal CGMs & $(2.4,25)$ & $(10,25)$ & $(1.2,24)$ & $(10,24)$ & $(2.5, 16)$ & $(5.8, 16)$ & $<1$ \\
        MW CGM & $(1,49)$ & $(12,49)$ & $(1,49)$ & $(12,49)$ & $(5.0,56)$ & $(16,55)$ & $(<1, 6.5)$
        \\ \hline
        \multicolumn{8}{c}{CR energy and power dissipation in the CGM for $\mhn=10^{12}$ M$_\odot$ at $z=0\ldots$}
        \\ \hline
        $E_{\rm CR,CGM}/|U_{\rm CGM}|$ & $(0.12,0.66)$ & $(0.12,0.27)$ & $(0.13,1.4)$ & $(0.13,0.31)$ &
        $(0.16,0.50)$ & $(0.16,0.31)$ & $(<0.01, 1.0)$
        \\ 
        $f_{\rm bolo}\dot{E}_{\rm st}/L_X$ & $(1.1, 1.9)$ & $(1.1, 1.8)$ & $(1.2, 1.9)$ & $(1.2, 1.8)$ & $(1.3, 1.9)$ & $(1.3, 1.8)$ & $(0.004, 0.56)$ \\
        \hline\hline
    \end{tabular}
    \caption{Summary of properties for the regions of parameter space not ruled out by observational constraints. In the list of parameter space regions, \textsc{TMax} is short for \textsc{TrappingMax} and similarly for \textsc{TMin}, and \textsc{FE} is short of \textsc{FreeEscape}. The different sections of the table report the range in $\alpha$ and $\log \mathcal{K}$ covered by each region, the percentage of CR energy that reaches the CGM of an $L_\star$ halo that has gone into each reservoir at $z=0$, the contribution of different emission components to the IGRB, ... Entries formatted as $(x,y)$ give the minimum and maximum value of each parameter over the parameter space region, while entries of the form $<x$ give an upper limit over the full region.
    \label{tab:cases}
    }
\end{table*}

\subsection{The cosmological cosmic ray energy budget}
\label{ssec:CR_budget}

\begin{figure}
    \centering
    \includegraphics[width=\columnwidth]{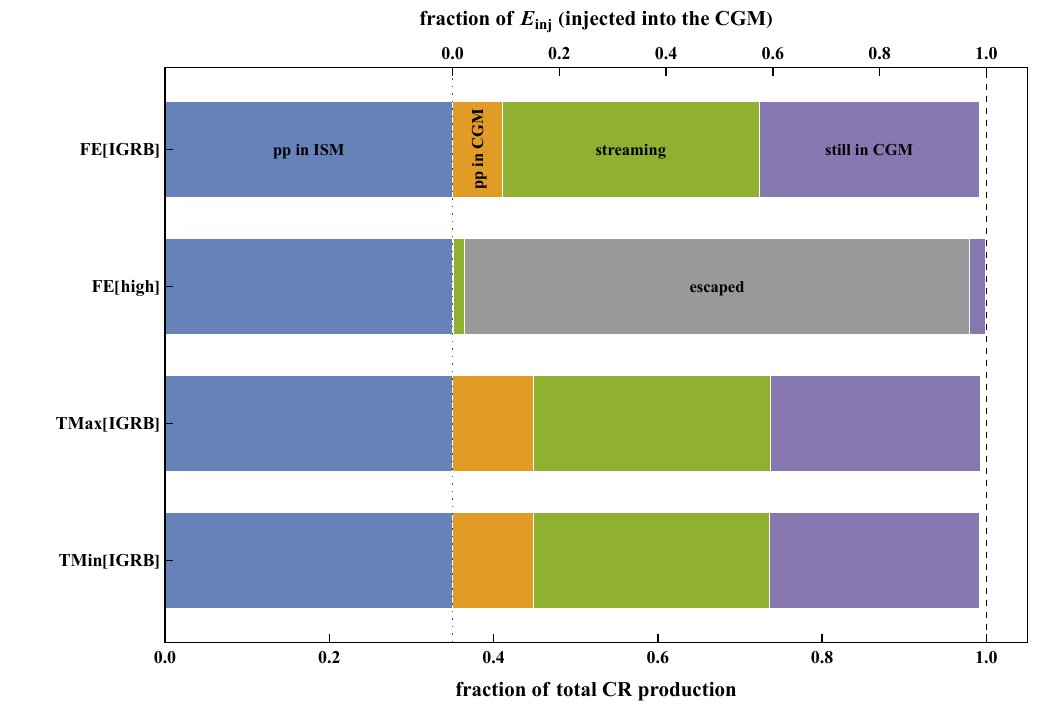}
    \caption{The fate of the CR energy produced over the star-formation history of a $10^{12}\,\msun$ halo, for four of the allowed regions of parameter space (see \autoref{tab:cases}). Each bar shows median energy going into each loss channel, as fractions of total CR production (bottom axis) and as fractions of the total CR energy that reaches the CGM rather than being lost in the disc (top axis). The reservoirs shown are hadronic losses in the ISM (first segment, ${\simeq}36$ per cent, independent of the CGM transport since it occurs before injection), hadronic losses in the CGM, streaming losses in the CGM, escape into the IGM, and remaining in the CGM to the present. Because each segment is a median taken component by component, total bar lengths do not sum exactly to unity.}
    \label{fig:barChartCRfateXpropFalseMasked}
\end{figure}

One output of our model is an answer to the budget question posed in the introduction: Where is all the energy that was injected into CRs as the Galaxy formed today? We answer this question in \autoref{tab:cases} and \autoref{fig:barChartCRfateXpropFalseMasked}, which shows the median result for how the CR energy produced over the star-formation history of an $L_\star$ halo is partitioned today for a subset of our defined parameter space regions. Some 36 per cent of the energy is lost hadronically in the ISM before the CRs reach the CGM, independently of the CGM model. Of the energy injected into the CGM, in all regions except \textsc{FreeEscape[high]}, two channels dominate: streaming losses, which heat the halo gas, take ${\sim}45$ per cent, and $27$--$55$ per cent is still resident in the CGM at $z\simeq0$; hadronic losses in the CGM take ${\sim}10$ per cent. By contrast, escape into the IGM is only signifcant  for \textsc{FreeEscape[high]}.

In absolute terms, for a $10^{12}\,\msun$ halo our fiducial model accelerates $\simeq4.1\times10^{58}$ erg in CR protons over all star formation since $z=8$, and delivers $2.6\times10^{58}$ erg ($65$ per cent) of this to the CGM after disc losses. Of this, in all cases except \textsc{FreeEscape[high]}, it retains
$E_{\rm CGM}\simeq1\times10^{58}$ erg in the CGM at $z\simeq0$. For scale, this reservoir is an order of magnitude larger than the CR proton content of the Fermi Bubbles in a hadronic scenario \citep{Crocker2011,Crocker2015,Tank2026}, ${\sim}10^{57}$ erg for an ambient density $n_{\rm H}\simeq10^{-3}\,{\rm cm^{-3}}$ and scaling as $n_{\rm H}^{-1}$. It amounts to $12$--$32$ per cent of the gravitational binding energy of the CGM (cf.~\autoref{ssec:cgm_budget}).

The persistence of significant CR energy in the CGM can be understood simply by examining timescales. The characteristic timescale for a CR injected at the halo centre to diffuse through its volume is determined by the time required for its three-dimensional random walk to cover a mean square distance $r_\star^2$; this timescale is
\begin{equation}
    t_{\rm diff} = \frac{\rvir^2}{6D} = \frac{1}{2\Kdiff}\frac{\rvir}{c_s}.
\end{equation}
Normalizing this to the Hubble time $t_{\rm H} = 1/H(z)$ (\autoref{ssec:halos}) and simplifying gives
\begin{equation}
    \frac{t_{\rm diff}}{t_{\rm H}} = \frac{1}{20\mathcal{K}} \Omega_{\rm m}(z)^{-1} \frac{v_{\rm vir}}{c_{\rm s}},
\end{equation}
with no explicit dependence on halo mass. For our adopted $\Omega_{\rm m}$ and $c_{\rm s}/v_{\rm vir}$ ratio (\aref{app:cgm_binding_energy}), our allowed range of $\Kdiff \approx 0.05-0.2$ corresponds to $t_{\rm diff}/t_{\rm H}\simeq 0.5-2$, with only a weak dependence on redshift. We therefore see that the observational constraints pick out a model in which CRs are able to traverse a halo roughly once per Hubble time. It is therefore not surprising that a significant fraction remain within it, even for \textsc{FreeEscape} boundary conditions.

\subsection{Where does the IGRB come from?}
\label{ssec:igrb_bbudget}

We can next use our model to identify where the IGRB comes from, decomposing it into contributions from extragalactic discs versus extragalactic CGMs versus the Milky Way's CGM, and in terms of the redshift and halo mass distribution of the extragalactic component.

We report the total contributions from both extragalactic CGM emission and from the Milky Way CGM for each of our allowed regions of parameter space in \autoref{tab:cases}; the disc contribution is fixed to $\simeq 60\%$ by the calibration we adopt in \aref{app:fcal_calibration} independent of CGM model parameters. We again see that the \textsc{FreeEscape[high]} region of parameter space is an outlier. In this case CRs escape from CGMs so quickly that there is negligible emission from them. Thus the $\simeq 40\%$ of the IGRB not directly supplied by  discs must come from $\gamma$-ray emission unconnected to star-forming galaxies. We return to the question of whether this is a realistic prospect in \autoref{ssec:high-K}, but for now we turn to the more interesting cases.

For cases where we choose parameters sufficient to reproduce the observed IGRB (\textsc{TrappingMax[IGRB]}, \textsc{TrappingMin[IGRB]}, and \textsc{FreeEscape[IGRB]}), we see that emission from extragalactic CGMs accounts for $\simeq 10-20\%$ of the IGRB; since extragalactic disc emission is $\simeq 60\%$, this means that CGM emission boosts the disc contribution by $\simeq 1/6-1/3$. The corresponding \textsc{All} cases simply extend the lower-edge of this range to somewhat lower values due to their lower values of $\alpha$, which place fewer baryons close to the galaxy where they are efficient targets for hadronic $\gamma$-ray production. We also see that the Milky Way halo makes a significant contribution of $\simeq 10-50\%$, with $\sim 25\%$ as a typical number. This is several times the $\sim 3-10\%$ estimated by \citet{Feldmann2013} for a Milky Way-like halo under a steady-state calorimetric model, and the reason for the difference is clear: for the range of parameters allowed by the observational constraints, a significant fraction of all the CRs ever produced over cosmological time are still present in the CGM (c.f.~\autoref{tab:cases}), and so the available energy budget to drive emission is significantly larger than the CR energy injected over a CGM crossing time as assumed in \citeauthor{Feldmann2013}'s steady-state model --- particularly since the star-formation rate of $L_\star$ halos was substantially higher at earlier cosmological epochs.

\begin{figure}
    \centering
    \includegraphics[width=0.5\textwidth]{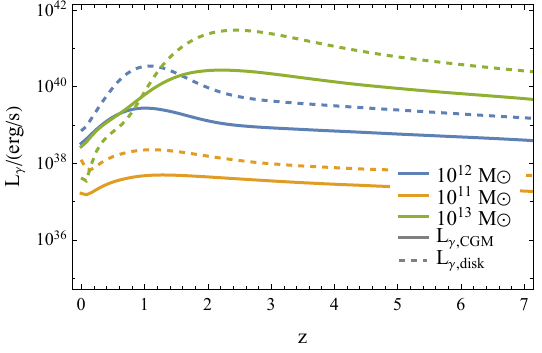}
    \caption{The $\gamma$-ray luminosity of the disc/ISM (dashed) and CGM (solid) of model halos with present-day masses $\mhn = 10^{11}$ (orange), $10^{12}$ (blue), and $10^{13}$ (green) M$_\odot$, for the example parameters $\alpha=2.3$ and $\log\Kdiff\simeq-0.81$, with \textsc{FreeEscape} boundaries.
    }
    \label{fig:plotLgammavszFE}
\end{figure}

We can see this effect more clearly by examining disc and halo $\gamma$-ray luminosity as a function of redshift and halo mass. We plot this in \autoref{fig:plotLgammavszFE} for the example point $\alpha=2.3$ and $\log\Kdiff\simeq -0.81$ with \textsc{FreeEscape} boundaries; this point lies close to the centre of the \textsc{FreeEscape[IGRB]} region, and the results for it are representative of those throughout the \textsc{IGRB} region for all boundary conditions. We see that disc emission (dashed lines) peaks at $z\sim 1$ for $\mhn=10^{12}$ M$_\odot$ halos and $z\sim 2-3$ for $\mhn=10^{13}$ M$_\odot$, roughly mirroring cosmic star formation histories for these halo masses, but dropping off more precipitously towards the present because both star formation rates (and thus CR injection) and $f_{\rm cal}$ are declining as the ISM becomes less dense. By contrast, CGM emission drops much more slowly because CRs reside in the CGM for long times, providing a delayed release of energy that was injected at earlier cosmic epochs when the star formation rate was higher. Thanks to this effect, at the present day the ratio of CGM $\gamma$-ray luminosity to disc $\gamma$-ray luminosity has risen to $\sim 1/2$ for $\mhn=10^{12}$ M$_\odot$ halos, from a minimum of $\sim 1/10$ at the peak of cosmic star formation. Larger halos are even more extreme: the CGM $\gamma$-ray luminosity of $\mhn=10^{13}$ M$_\odot$ \textit{exceeds} the disc luminosity by almost an order of magnitude at $z\simeq 0$, because star formation, and so CR injection, has effectively ceased while the CRs injected into the CGM at earlier epochs continue to radiate.

The pattern of evolution shown in \autoref{fig:plotLgammavszFE} also dictates the distribution of redshift and halo mass that contributes to the IGRB. Both the disc and the extragalactic CGM contributions peak in Milky Way-scale and larger halos ($M_{\rm h,0}\sim10^{12}$--$10^{13}\,\msun$) near the epoch of most intense cosmic star formation ($z\sim1$), with the CGM term weighted to somewhat more massive halos. The background is therefore dominated by $L_\star$-type galaxies near the cosmic star-formation peak, rather than by dwarf halos or by groups and clusters.

\subsection{The CR contribution to the CGM pressure support}
\label{ssec:cgm_budget}

\begin{figure}
    \centering
    \includegraphics[width=\columnwidth]{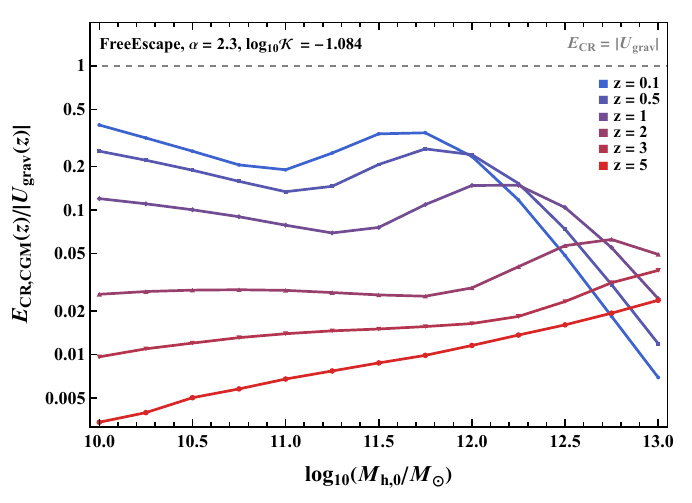}
    \caption{Ratio of CR energy to gas gravitational binding energy in the CGM, $E_{\rm CR,CGM}/|U_{\rm CGM}|$, as a function of present-day halo mass $\mhn$ at six redshift slices from $z=0.1$ (blue) to $z=5$ (red), for \textsc{FreeEscape} at $(\alpha,\log_{10}\Kdiff)=(2.3,-1.084)$; the dashed horizontal line marks $E_{\rm CR,CGM}=|U_{\rm CGM}|$.}
    \label{fig:plotEddEvol1}
\end{figure}

Cosmic rays can matter to the CGM in two logically distinct ways. The first is mechanical: if the retained CR energy is comparable to the gravitational binding energy of the halo gas, CR pressure contributes to its support and the gas cannot be in purely thermal hydrostatic equilibrium. The second is thermal: if the CR heating power is comparable to the halo's radiative losses, cosmic rays can hold the gas against cooling. We examine these in turn.

To characterise the mechanical importance of CRs we compute the ratio $E_{\rm CR,CGM} / |U_{\rm CGM}|$, where $E_{\rm CR,CGM}$ is the total energy of CRs in the CGM integrated over all radii and $U_{\rm CGM}$ is the CGM binding energy (\aref{app:cgm_binding_energy}). We report the range of this ratio for $\mhn=10^{12}$ M$_\odot$ halos at $z=0$ over our allowed parameter regions in \autoref{tab:cases}. As with all other results in this section, in the \textsc{FreeEscape[High]} case the energy is small for most parameters in this region, although there is a small corner of parameter space where it can be relatively large. However, we will focus more on the other possible solution where CRs are not assumed to escape the halo rapidly. We plot $E_{\rm CR,CGM} / |U_{\rm CGM}|$ as a function of $\mhn$ and $z$ for a representative set of parameters in this region -- $\alpha = 2.3$, $\log\Kdiff=-1.084$, with \textsc{FreeEscape} boundaries -- in \autoref{fig:plotEddEvol1}.

The figure shows that the $E_{\rm CR,CGM} / |U_{\rm CGM}|$ ratio grows steadily with time for $\mhn\lesssim10^{12}\,\msun$, reflecting the continuous accumulation of CR energy across the star-formation history; at $10^{12}\,\msun$ it is roughly constant at late times, and above that mass it has declined from an earlier peak. By $z\simeq0$, Milky Way-mass halos reach $E_{\rm CR,CGM}/|U_{\rm CGM}|\simeq0.24$, and $12$--$32$ per cent across the \textsc{IGRB} regions of parameter space: CR pressure support is energetically significant in $L_\star$ CGMs at low redshift. halos substantially more massive than $L_\star$ remain well below the CR Eddington condition even after a full Hubble time of accumulation. Probes of the large-scale pressure profile --- X-ray surface brightness, the thermal Sunyaev--Zel'dovich signal \citep{Kim2022}, and absorption-line diagnostics \citep{Werk2014} --- should in principle be able to confirm this prediction \citep{Ji2020, Quataert2025}; indeed, O\,\textsc{viii} absorption has already been used to bound the CR-to-thermal pressure ratio in the Milky Way CGM \citep{Roy2022}.

\begin{figure}
    \centering
    \includegraphics[width=\columnwidth]{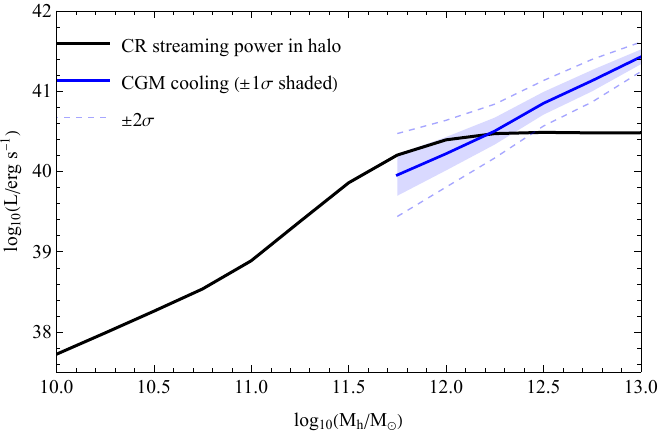}
    \caption{Cosmic-ray streaming power delivered to the halo (black) compared with the radiative output of the CGM derived from \textit{eRosita} stacks \citep{Zhang2024eROSITAII} corrected by dividing by the bolometric factor $f_{\rm bolo} = 1/3$ (blue solid line, with the $\pm1\sigma_{\rm dex}$ band shaded and $\pm2\sigma_{\rm dex}$ shown as faint dashed lines), both at $z\simeq0$ as a function of present-day halo mass calculated for \textsc{FreeEscape}, $\alpha$ = 2.3, and $\Kdiff \simeq$ 0.15. The cooling curve is drawn only above $M_{\rm h}\simeq10^{11.75}\,\msun$, the lowest halo mass probed by observations.}
    \label{fig:crVsCool}
\end{figure}

Turning to the thermal importance of CRs, in \autoref{fig:crVsCool} we compare the streaming power delivered to the CGM by CRs at $z=0$ as a function of halo mass to the observed X-ray luminosity as a function of halo mass (c.f.~\autoref{ssec:erosita}) divided by our adopted bolometric correction factor $f_{\rm bolo}=1/3$ (c.f.~\autoref{sssec:model_prediction_xray}); this plot uses the same set of example parameters as in \autoref{fig:plotEddEvol1}. We also report the ratio $f_{\rm bolo} \dot{E}_{\rm st}/L_{\rm X,CGM}$ for $\mhn=10^{12}$ M$_\odot$ at $z=0$ in \autoref{tab:cases}. We see that CR streaming dissipation is comparable to the observed X-ray luminosity (after applying $f_{\rm bolo}$) for halos close to $\mhn\approx 10^{12}$ M$_\odot$, but not at higher masses, a pattern similar to that seen in our comparison of CR energy density to CGM binding energy.

That both $E_{\rm CR,CGM} / |U_{\rm CGM}|$ and $f_{\rm bolo} \dot{E}_{\rm st}/L_{\rm X,CGM}$ fall above $L_\star$ --- the energy ratio from a local maximum just below it, the power comparison crossing unity at it --- is not a coincidence. Instead, this behaviour is a result of the same fundamental budget considerations that make stellar feedback increasingly ineffective above $L_\star$. The CR energy released per unit mass of stars formed, $\eta_c\simeq10^{48}\,{\rm erg}\,\msun^{-1}=5.0\times10^{14}\,{\rm cm^{2}\,s^{-2}}$, is to be compared with the specific binding energy of the halo, $v_c^{2}=GM_{\rm h}/\rvir$, which is $1.4\times10^{14}\,{\rm cm^{2}\,s^{-2}}$ at $10^{12}\,\msun$; the ratio is $\simeq3.6$. Folding in the efficiency with which halo baryons are converted into stars, $\varepsilon\simeq0.17$ at this mass, gives a CR energy per unit halo gas mass of $\simeq0.6\,v_c^{2}$ --- of order unity. Feedback regulates where the energy returned per unit mass is comparable to the binding energy per unit mass, and cosmic rays carry a fixed fraction of the supernova energy, so the CR budget inherits that. Beyond $\mh \simeq 10^{12}$ M$_\odot$, however, halo binding energy rises as roughly $\mh^{5/3}$ (\aref{app:cgm_binding_energy}) while X-ray luminosity continues to rise as roughly $\mhn^{1.3}$ (cf.~\autoref{fig:crVsCool}), and in this halo mass range star formation efficiency is either flat or falling, so the available CR energy budget becomes increasingly small in comparison.

\subsection{CR transport coefficients in the CGM}
\label{ssec:cgm_transport}

A final implication of our results is that -- again excluding the outlier \textsc{FreeEscape[High]} region -- the various observational constraints available pin $\Kdiff$ to a relatively narrow range from $\approx 0.05-0.2$. In dimensional rather than dimensionless terms, for a Milky Way-mass halo at $z=0$, this corresponds to a diffusion coefficient $\approx 2-7 \times 10^{29} (\mhn/10^{12}\,\msun)^{0.66}$ cm$^2$ s$^{-1}$ at $z = 0$. It is therefore interesting to compare this range to published observational estimates of CR transport rates in the CGM.

This value is consistent with the cluster lower limits of \citet{Wiener2019} and \citet{Habegger2026} and brackets the group-scale value of \citet{Yamasaki2024} for the hosts of odd radio circles interpreted as virial shocks, provided that we correctly include the scaling with characteristic halo mass, which differs for each of these observations. Comparison with the results of \citet{Butsky2023} is more challenging because for simplicity we assume a single $\Kdiff$ throughout a halo, while they attempt to estimate the radial variation of the diffusion coefficient. Their lower limits on the effective transport rate around star-forming COS-Halos galaxies \emph{rise} with radius over $r\simeq0.06$--$0.6\,r_{\rm vir}$, so our physical diffusion coefficient sits within their inner-CGM limits and a factor ${\sim}3$ below their outer-CGM floor.

The comparison to \citet{Butsky2023} also points to a limitation of our model, which is that that the radial dependence of the diffusion coefficient is essentially unconstrained in our model, due to a near-exact degeneracy. An outwardly rising $\Kdiff(R)$ would, with a flatter density profile,  yield nearly the same inner-CGM CR density and $\gamma$-ray output as a steeper gas profile at constant $\Kdiff$, so the range of slopes where our allowed models sit ($\alpha\approx1.8$--$2.9$ for the \textsc{IGRB} cases) should be read as an effective parameter absorbing that degeneracy rather than as a direct measurement of the CGM density slope.

We defer further discussion of the implications of our findings for $\Kdiff$ for the nature of CR transport in the CGM, and the physical origins of the transport rate, to Paper III.

\subsection{Are high $\Kdiff$ models plausible?}
\label{ssec:high-K}

We have seen throughout our discussion of implications that the \textsc{FreeEscape[High]} region of parameter space, in which CRs rapidly escape into the IGM and thus are dynamically and energetically irrelevant to the CGM, leads to conclusions that differ markedly from the other allowed regions of parameter space. It is therefore worth investigating this region further.

One potential concern regarding this region of parameter space is that adopting it requires at least a minor conspiracy to reproduce the observed IGRB. In this scenario, the ${\gtrsim}20$ per cent of the IGRB that the CGM supplies when $\Kdiff$ falls into the lower allowed regions would have to come from elsewhere, and whatever supplies it must combine with a pionic disc spectrum, $\alpha_p^{\rm disk}\simeq2.35$, to reproduce the observed single-power-law shape, $\Gamma\simeq2.32$ over 1--10\,GeV (\aref{app:fcal_calibration}, \autoref{fig:IGRBspecFit}). Since the disc component already carries that shape, so must the remainder; this constraint immediately disfavours some plausible IGRB contributors, including millisecond pulsars and dark-matter annihilation. Sources that are spectrally similar to star-forming galaxies in the 1--10,GeV band such as faint blazars and misaligned AGN \citep[e.g.,][]{Ajello2015} could in principle supply the remainder, but since they draw on an unrelated power source -- accretion onto supermassive black holes -- for them to make up the balance left after the star-forming disc contribution would require the two to agree to within a factor of a few. That is possible, but seems a somewhat uncomfortable coincidence (albeit one that galaxy--black hole co-evolution might partly explain). The circumgalactic emission of the same star-forming population, by contrast, carries an almost identical and slightly harder spectrum, $\alpha_p^{\rm CGM}\simeq2.2$, and supplies the remainder with no coincidences required.

One could in principle avoid this problem by positing that star formation supplies only a minor share of the IGRB, but this option is harder still to sustain. Discs could radiate less only if their calorimetry were well below what the \congr{} treatment gives (i.e., lower $f_{\rm cal}$), or if the overall luminosity per unit star formation were lower (lower $\eta_c$) -- and both these options run into obvious problems. The \congr{} value of $f_{\rm cal}$ reproduces the broadband non-thermal spectra of local star-forming galaxies at the factor-of-two level \citep{Roth2021,Roth2023} and it is hard to see how to significantly reduce $f_{\rm cal}$ without breaking this. Moreover, lower $f_{\rm cal}$ would add to $E_{\rm CGM}$, pushing $L_\star$ halos towards the non-hydrostatic regime. Lowering $\eta_c$ runs into a similar problem: applied to the Galaxy's measured star-formation rate of $1.65\pm0.19\,\msun\,{\rm yr^{-1}}$ \citep{Licquia2015}, our fiducial $\eta_c$ implies predicts a luminosity $5.8\times10^{40}\,{\rm erg\,s^{-1}}$ in CR protons, already at the low end of the
$(6.0$--$7.4)\times10^{40}\,{\rm erg\,s^{-1}}$ inferred from local CR data by \citet{Strong2010}.

We are therefore left with the conclusion that the high-$\Kdiff$ region of parameter space is not directly ruled out, but that it requires the existence of a second class of source that contributes to the IGRB that is, coincidentally, very similar in both total luminosity and spectral shape to the emission from the discs of star-forming galaxies.

\section{Summary, conclusions, and prospects}
\label{sec:conclusion}

In this paper we construct a simple, phenomenological model for the transport of CRs through the CGM. We then ask under what circumstances this model is consistent with observational constraints arising on $\gamma$-ray and X-ray emission from the CGM, and with the requirement that the CR pressure in the CGM not be so large as to render it dynamically unstable. Our primary conclusions are as follows:

\begin{itemize}
\item All the observational constraints are satisfied in a region of parameter space where CR transport rates in the CGM are moderate. Quantitatively, for a diffusion coefficient parametrised as $D_{\rm eff}=\Kdiff\,\tfrac13 r_{\rm vir}c_s$, all the constraints are satisfied for $\Kdiff \simeq 0.05-0.2$.
In this region, the diffusion time across an $L_\star$ halo is of order a Hubble time, so CRs accumulate over galaxies’ star formation histories and are dynamically significant, though not so dominant as to render hydrostatic balance in the CGM impossible.
\item For this transport rate, hadronic emission from star-forming galaxies fully explains the observed isotropic $\gamma$-ray background (IGRB), as long as the CGM gas density distribution obeys $\rho \propto r^{-\alpha}$ with $\alpha\simeq1.8$--$2.9$. In this regime roughly 60\% of the IGRB is produced by extragalactic discs, $\simeq 10-16\%$ from extragalactic CGMs, and $\simeq 24-28\%$ from the Milky Way's own CGM.

\item In this same region of parameter space, with no tuning of parameters, heating of CGM gas by CR streaming losses also naturally explain the observed X-ray luminosities of $L_\star$ galaxy halos. 

\item The only other scenario that is consistent with all the observational constraints is one where CRs transit through the CGM rapidly, $\Kdiff \gtrsim 1$, and 
escape freely into the IGM, with  CR pressure itself
plausibly preventing the formation of a confining virial shock.
In this scenario, CR pressure and heating in the CGM are both negligible, and explaining the IGRB then requires positing the existence of 
a second source of $\gamma$ rays whose spectrum and luminosity closely match those of star-forming discs to supply the remaining $\simeq40$ \% of the IGRB. While not a definitive argument, we argue that this favours the first scenario on the grounds of simplicity.
\end{itemize}

Taken together, our numerical experiments show that, even though CRs in the CGM are at present extremely difficult to detect, current observations do set meaningful constraints on their properties. These constraints favour a scenario where the CR diffusion time through the halos of $L_\star$ galaxies is of order the Hubble time, and as a result CRs accumulate in the CGM, contributing meaningfully to its dynamics, heating, and $\gamma$-ray emission. 

While this result is suggestive, significant territory remains to be explored. In our simple model we have not yet considered the effects of varying either the efficiency with which star formation accelerates CRs or the magnetisation of the CGM. Nor have we fully explored all the available observational avenues for constraining the fate of CRs in the CGM, or the implications of our empirical constraints for the physical nature of CR transport. Pulling out these threads will be the work of the next two papers in this series: Paper~II will draw out the observational consequences of our model including halo-mass-resolved $\gamma$-ray stacking, radio and X-ray signatures, while Paper~III will ask whether self-confinement and damping can physically produce $\Kdiff\simeq0.1$ and how that depends on CGM magnetisation.

\section*{Acknowledgements}
RMC thanks Felix Aharonian, Lucia Armillotta, Yuval Birnboim, Yakov Faerman, Volker Heesen, Cassi Lochhaas, Jens Niemeyer, Christoph Pfrommer, Martin Pohl, Gabriele Ponti, Matt Roth, Andy Sullivan, Andrew Taylor, Siyao Xu, and Ellen Zweibel for helpful conversations about the physics of CRs and/or the CGM.
RMC acknowledges further useful conversations with participants in the 
Aspen Center for Physics meeting `Cosmic Ray Feedback in Galaxies and Galaxy Clusters'. The Aspen Center is supported
by National Science Foundation grant PHY-2210452.
RMC also thanks
Chris Stevens and Arash Tavassoli for \textsc{MATHEMATICA} coding advice and assistance, and Kate Latimer \& Lyndall Kennedy and Dav Taylor \& Michelle Styles for hospitality and a quiet place to work. The authors acknowledge the use of \textsc{Claude Code} in preparing this text and some of the
\textsc{MATHEMATICA} visualisation and data processing code; the core PDE algorithm was developed by the authors.
The authors thank
Phil Hopkins for provision of data and Matt Roth
for providing \congr{} outputs and for early input to this project. The authors acknowledge support from the Australian Research Council through award DP260100433.

\section*{Data Availability}
The cosmic-ray transport solutions, the post-processed model grids, and the \textsc{Mathematica} code used to generate the figures and derived quantities in this article will be shared on reasonable request to the authors. The observational data underlying the comparisons -- the \textit{Fermi}-LAT $\gamma$-ray measurements, the eROSITA X-ray surface-brightness profiles, and the IGRB and outer-Galaxy ring-emissivity compilations -- are all publicly available from the cited references.

\bibliographystyle{mnras}
\bibliography{HaloCRs}

\appendix

\section{Halo density structure, virial temperature, and binding energy}
\label{app:cgm_binding_energy}

We assume that dark matter in halos is distributed following an NFW profile \citep{Navarro1997}
with scale radius $r_s$, and write $\mathcal{C}\equiv\rvir/r_s$, so that
$\rho_{\rm DM}\propto1/[(\mathcal{C}r/\rvir)(1+\mathcal{C}r/\rvir)^2]$. Our virial radius
$\rvir=r_{200\rm m}$ encloses a mean density $200\bar\rho_{\rm m}(z)=200\,\Omega_{\rm m}(z)\rho_{\rm crit}(z)$,
so $\mathcal{C}$ is not the concentration $c_{200\rm c}=r_{200\rm c}/r_s$ (with $r_{200\rm c}$
enclosing $200\rho_{\rm crit}$) given by the relation of \citet[their section 4.3 and appendix C]{Ludlow2016}.
The two are related by
\begin{equation}
\frac{g(\mathcal{C})}{\mathcal{C}^{3}}=\Omega_{\rm m}(z)\,\frac{g(c_{200\rm c})}{c_{200\rm c}^{3}},
\end{equation}
with $g$ as in \autoref{eq:curlyg}. The relation of \citet{Ludlow2016} must be evaluated at $M_{200\rm c}=\mh\,g(c_{200\rm c})/g(\mathcal{C})$ (iterated to convergence). For an $L_\star$ halo at $z=0$ this gives $c_{200\rm c}=8.9$ and
$\mathcal{C}=14.2$ ($M_{200\rm c}=0.78\,\mh$, $r_{200\rm m}=1.6\,r_{200\rm c}$); the two coincide as $\Omega_{\rm m}(z)\to1$.

The fraction of the total halo mass $\mh$ at radii $<r/\rvir$ is
\begin{equation}
    f_M = \frac{\ln (1 + \mathcal{C}r/\rvir) - 1 + (1 + \mathcal{C}r/\rvir)^{-1}}{\ln(1+\mathcal{C}) - 1 + (1 + \mathcal{C})^{-1}}.
\end{equation}
The mass $M_{500c}$ (required for the X-ray analysis in \autoref{ssec:erosita}) is that within the
radius $r_{500c}$ inside which the mean density is $500\rho_{\rm crit}(z)$, whereas our halo mass
$\mh=M_{200m}$ is that within $\rvir$, inside which the mean density is $200\,\Omega_{\rm m}(z)\rho_{\rm crit}(z)$.
In terms of $y\equiv r/\rvir$ the mean enclosed density is $200\bar\rho_{\rm m}f_M(y)/y^3$, so
$r_{500c}=y_{500}\rvir$ with $y_{500}$ the root of $f_M(y_{500})/y_{500}^{3}=5/[2\Omega_{\rm m}(z)]$
($=7.9$ at $z=0$), and $M_{500c}=f_M(y_{500})\,\mh$. For an $L_\star$ halo at $z=0$ ($\mathcal{C}=14.2$)
this gives $r_{500c}\simeq0.43\,\rvir$ and $M_{500c}=0.61\,\mh$.

The virial temperature (\autoref{ssec:trans_eq}) is set by the gravitational self-energy of the
halo truncated at $\rvir$, which is dominated by the dark matter, and given by \citep{Mo1998} $U_{\rm grav} = -f_\mathcal{C}\,G\mh^2/\rvir$, with
\begin{equation}
    f_\mathcal{C} = \frac{\mathcal{C}}{2}\,
    \frac{1 - (1+\mathcal{C})^{-2} - 2\ln(1+\mathcal{C})/(1+\mathcal{C})}{g(\mathcal{C})^2},
\end{equation}
and
\begin{equation}
    g(\mathcal{C})=\ln(1+\mathcal{C})-\frac{\mathcal{C}}{1+\mathcal{C}};
\label{eq:curlyg}
\end{equation}
$f_\mathcal{C}=1.42$ for $\mathcal{C}=14.2$. 
$U_{\rm grav}$, and hence $T_{\rm vir}$ and $c_s$, are evaluated with $\mathcal{C}=c_{200\rm m}$ from the relation above ($f_\mathcal{C}=1.42$ for an $L_\star$ halo at $z=0$), and the same $\mathcal{C}$ sets the NFW gravity $g(r)$ in $\eta_P$ and $U_{\rm CGM}$.

The dark matter binding energy is distinct from,
and much larger than, the binding energy of the CGM gas in the halo potential, $U_{\rm CGM}$, which we write as $U_{\rm CGM} = \mathcal{U}\,G\mh\mcgm/\rvir$, where $\mathcal{U}$ is a dimensionless function. To compute $\mathcal{U}$, we first note that for an NFW halo truncated at $\rvir$, the dimensionless gravitational potential $\hat{\phi} = \phi/(G\mh/\rvir)$ at dimensionless radius $R = r/\rvir$ is
\begin{equation}
    \hat{\phi} =
    \begin{cases}
        -1+\dfrac{\ln(1+\mathcal{C})}{g(\mathcal{C})}-\dfrac{\ln(1+\mathcal{C} R)}{R\,g(\mathcal{C})}, & R \leq 1, \\
        -1/R, & R > 1,
    \end{cases}
\end{equation}
where $g(\mathcal{C})$ is defined above (\autoref{eq:curlyg})
and the gravitational potential energy of the CGM  within $R$ is $U_\mathrm{CGM} = \rvir^3 \int_{R_0}^{R} 4\pi R^2 \rho_\mathrm{g,in} \phi \, dR$, where $\rho_{g,\rm in}$ is given by \autoref{eq:cored_profile}.

The choice of the outer radius to which we integrate to evaluate $U_{\rm CGM}$ depends on the choice of boundary condition (c.f.~\autoref{sssec:cgm_model}). For \textsc{TrappingMax} the full CGM mass $\mcgm$ is located between $R = R_0$ and 1, and thus the outer boundary is at $R = 1$. For \textsc{TrappingMin} part of the CGM mass resides outside the virial shock, but we choose to count towards our binding energy only the material inside the virial shock, since the CRs are assumed to be confined within $\rvir$ and thus to make a fair comparison we only wish to include the binding energy of the material cospatial with them. Thus we again choose $R = 1$ as the outer boundary. Finally, for the \textsc{FreeEscape} case the CGM and CR distributions both continue out to the closure radius $R_\mathrm{clos}$, so we take this as our outer boundary.

With these choices, we now define the dimensionless CGM gas density profile by $\hat{\rho} = (R^2 + R_0^2)^{-\alpha/2}$, and we can write down our expressions for the CGM binding energy, and thus for the dimensionless function $\mathcal{U}$. For \textsc{TrappingMin} and \textsc{TrappingMax} we have
\begin{equation}
    \mathcal{U}_{\rm trapping}=
    -f_\mathrm{in} \frac{\int_{R_0}^{1} R^2\hat\rho\,\left(1+\hat\phi\right)\, dR}
    {\int_{R_0}^{1} R^2\hat\rho\, dR},
\end{equation}
where $f_\mathrm{in}$ is the fraction of CGM material interior to $\rvir$ ($=1$ for \textsc{TrappingMax}, and determined by the numerical solution for $\rho_0$ for \textsc{TrappingMin} -- c.f.~\autoref{sssec:cgm_model}); for the \textsc{FreeEscape} case we have instead
\begin{equation}
    \mathcal{U}_\mathrm{free-escape}
    =
    -\frac{\int_{R_0}^{R_\mathrm{clos}} R^2\hat\rho \left(R_\mathrm{clos}^{-1} + \hat{\phi}\right) \, dR}{\int_{R_0}^{R_\mathrm{clos}} R^2\hat\rho\, dR},
\label{eq:Ufnctl}
\end{equation}
These dimensionless functions are straightforward to evaluate numerically; for the trapping modes, $\mathcal{U}_{\rm shock}/f_{\rm in}$ equals the \textsc{FreeEscape} function evaluated at $R_{\rm clos} = 1$.

\section{Calibration of the effective energy-integrated calorimetry fraction}
\label{app:fcal_calibration}

Our galaxy-by-galaxy calorimetry is the kernel density estimate built directly from \congr{}'s own $f_\mathrm{cal}$ records (cf.~\autoref{sssec:CR_injection}). That estimate carries no energy dependence, and so cannot reproduce the \emph{spectrum} of the summed disc emission along the light cone. The parameters $\delta\alpha$ and $q_{\rm cal}$ repair precisely this: $\delta\alpha$ sets the spectral weighting with which we collapse $f_\mathrm{cal}(E_{\rm p})$ to a single number, and $q_{\rm cal}$ its normalisation. No equivalent tweak is needed for the CGM, whose emission emerges with essentially the spectrum at injection.

To fix $\delta\alpha$ and $q_{\rm cal}$, we first generate 4D data sets $\{\log M_\star,\log \dot{M}_\star,\log r_{\rm d}, f_\mathrm{cal}(\delta\alpha)\}$ as in \autoref{sssec:CR_injection}, but over a fine grid of spectral tilts $\delta\alpha$ (no grid in $q_{\rm cal}$ is needed, since it is a constant multiplicative factor), and apply the same \textsc{Mathematica} kernel density estimation to each. This yields $f_\mathrm{cal}$ for any halo of mass $\mh$ at redshift $z$ and any $q_{\rm cal}$.

We use this prediction to calculate the contribution to the IGRB from galactic discs. We assume that a halo of mass $\mh$ at redshift $z$ hosts a disc with total $\gamma$-ray luminosity $L_{\gamma,\rm disc} = \frac{1}{3} f_\mathrm{cal} \eta_c \mdotstar$, with $f_{\rm cal}$ and $\mdotstar$ evaluated from $\mh$ and $z$, and specific luminosity $dL_{\gamma,\rm disc}/dE_\gamma = L_{\gamma,\rm disc}\, s_\gamma(E_\gamma)$, where $s_\gamma(E_\gamma)$ is a spectral shape function with $\int s_\gamma(E_\gamma) \, dE_\gamma = 1$. We compute $s_\gamma(E_\gamma)$ for a proton population with momentum powerlaw index \(\alpha_{p,0} + \delta\alpha\), where we adopt $\alpha_{p,0} = 2.2$ as the spectrum of protons at acceleration, using the inclusive \(pp\to\pi^0\to\gamma\gamma\)
differential cross-section parametrisation of \citet{Kamae2006} at low proton energies and \citet{Kelner2006} (with their inelastic \(pp\) cross-section) at high energies.

The total disc contribution to the IGRB flux seen from Earth is
\begin{align}
    \frac{dF_{\gamma,{\rm disc}}}{dE_{\gamma,0}}
    &=
    \int dz
    \int dM_{\rm h,0}\,
    \frac{dn}{dM_{\rm h}}\frac{d\mh}{d\mhn}\,
    \frac{c}{H(z)(1+z)}
    \nonumber\\
    &\qquad\times
    L_{\gamma,{\rm disc}} s_\gamma\!\left(
    (1+z)E_{\gamma,0}
    \right)
    ,
    \label{eq:disk_lightcone_diff}
\end{align}
where $E_{\gamma,0}$ is the photon energy observed at Earth and the remaining symbols are as in \autoref{eq:igrb_lightcone}; the kernel here is $c/[H(z)(1+z)]$, rather than $c/[H(z)(1+z)^2]$, because $dF_{\gamma,{\rm disc}}/dE_{\gamma,0}$ is a spectral flux density. All quantities that depend on the halo mass $\mh$ are evaluated for halos of mass $\mh(\mhn,z)$. Integrating over the $1$--$10$ GeV band, $\int s_\gamma\!\left((1+z)E_{\gamma,0}\right)dE_{\gamma,0}=f_{\rm bolo}/(1+z)$ turns $c/[H(z)(1+z)]$ into the $c/[H(z)(1+z)^2]$ of \autoref{eq:igrb_lightcone}.

We then calibrate our choice of $\delta\alpha$ and $q_{\rm cal}$ by optimising them so that the value of $dF_{\gamma,\rm disc}/dE_{\gamma,0}$ computed from \autoref{eq:disk_lightcone_diff} reproduces as closely as possible the hadronic ($\pi^0$) disc emission computed directly from \citet{Roth2021}, rescaled to match the slightly differing assumptions we make here compared to \citeauthor{Roth2021}. These differences are as follows. First, \citet{Roth2021} adopt $1.32\times10^{-2}$ supernovae per $\msun$ of stars formed, equivalent to $10^{50}$ erg per $75.7\,\msun$, compared to our $10^{50}$ erg per $90\,\msun$ (\autoref{eq:cr_injection_efficiency}); since $98\%$ of the injected energy lies above the pion production threshold, we compensate for this by reducing the \citet{Roth2021} prediction by a factor $1.17$. Second, as pointed out by \citet{Chen2025}, there is a mismatch between the \citet{Salpeter55a} IMF assumed by \citet{Madau2014} in their calculation of the star formation history of the Universe and the \citet{Chabrier2005} IMF used to infer star formation rates in the CANDELS sample and assumed here; we correct for this by multiplying the \citeauthor{Roth2021} result by a factor of $0.63$. Third, the star-formation history implied by our adopted model in \autoref{ssec:halos} is smaller than the \citet{Madau2014} one (corrected to a \citeauthor{Chabrier2005} IMF) by a factor of $0.90$ (\autoref{fig:cosmicvalidation}). Combining these three factors, our calibration target is the \citet{Roth2021} $\pi^0$ prediction multiplied by $0.48$. Note that the galaxy-by-galaxy calorimetry is \emph{not} rescaled: our model galaxies are less calorimetric than CANDELS galaxies at the same epoch, and that difference is a prediction of our halo model, already present in \autoref{eq:disk_lightcone_diff}.

With this rescaling in place, we then choose $\delta\alpha$ and $q_{\rm cal}$ as the values that minimise the residual
\begin{eqnarray}
    \lefteqn{
    R(\delta\alpha,q_{\rm cal}) = \frac{1}{N} \sum_{i=1}^N 
    }
    \nonumber \\
    & & \left[
    \ln \left(E_{\gamma,0}\frac{dF_{\gamma,{\rm disc}}}{dE_{\gamma,0}}\right)_{{\rm cal},i} -   \ln \left(E_{\gamma,0}\frac{dF_{\gamma,{\rm disc}}}{dE_{\gamma,0}}\right)_{{\rm CONG},i} \right]^2, \qquad
\end{eqnarray}
where $(dF_{\gamma,{\rm disc}}/dE_{\gamma,0})_{{\rm cal},i}$ is our calibration expression given by \autoref{eq:disk_lightcone_diff} evaluated at photon energy $E_{\gamma,0,i}$ and for a particular choice of $\delta\alpha$ and $q_{\rm cal}$, $(dF_{\gamma,{\rm disc}}/dE_{\gamma,0})_{{\rm CONG},i}$ is the \congr{} prediction at the same photon energy, rescaled as above, and the sum is evaluated over the \congr{} output energy grid from 1 - 10 GeV. We restrict the fit to this range because below 1 GeV our treatment of the $\pi^0$ kinematic cutoff is simpler than the one implemented in \congr{}, and above 10 GeV $\gamma\gamma$ opacity becomes significant and we do not model it; within the window that opacity is below $2\%$.

\begin{figure}
\centering
\includegraphics[width=0.8\columnwidth]{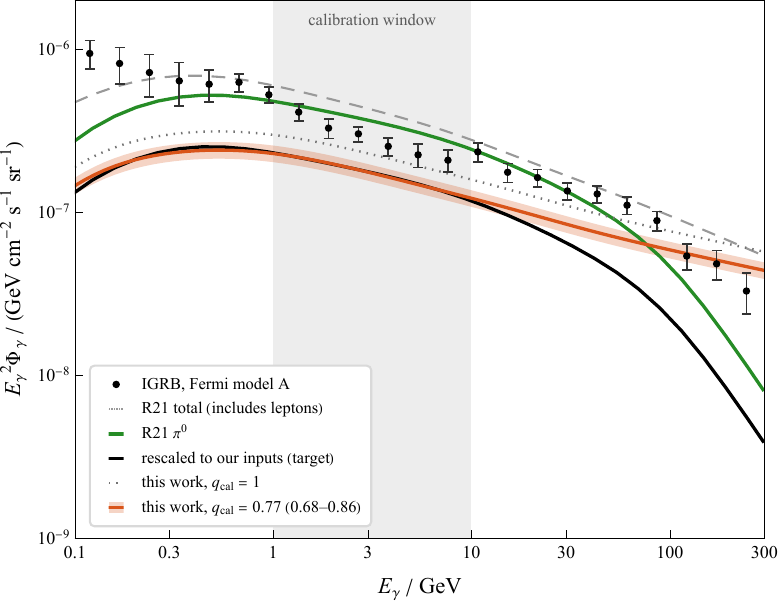}
\caption{
Calibration of the effective disc calorimetry. Points: the Fermi model-A IGRB \citep{Ackermann2015}. Green: the \congr{} hadronic ($\pi^0$) disc light cone \citep{Roth2021}; grey dashed, their total, which includes the leptonic components we do not model. Black: their $\pi^0$ prediction rescaled to our inputs — our cosmic-ray yield per unit stellar mass formed, the initial-mass-function normalisation of the cosmic star-formation correction \citep{Chen2025}, and our star-formation history. Orange: our effective, spectrally tilted treatment with $\delta\alpha=0.15$ at the adopted $q_{\rm cal}=0.77$, the shaded band showing the calibration range $0.68$–$0.86$; dotted grey, the same at $q_{\rm cal}=1$. The vertical shaded band marks the $1$–$10$ GeV window in which $\delta\alpha$ and $q_{\rm cal}$ are fitted; there the residual is $0.005$ dex rms, or $1.2\%$.
\label{fig:IGRBspecFit}
}
\end{figure}

\begin{figure}
    \centering
    \includegraphics[width=0.7\columnwidth]{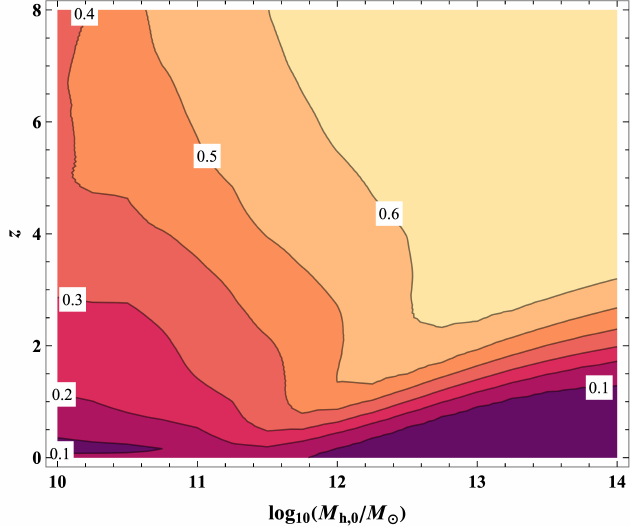}
    \caption{
    Calibrated effective CR proton calorimetry fraction $f_{\rm cal}$ for  $\delta\alpha = 0.15$ and $q_{\rm cal} = 0.77$ as a function of present-day halo mass and redshift.}
    \label{fig:plotfcal}
\end{figure}

The residual is minimised at 
$\delta\alpha = 0.15 \pm 0.02$ (where the uncertainty is the range over which the fit residual doubles)
and $q_{\rm cal} = 0.77$, where \autoref{eq:disk_lightcone_diff} reproduces the target to $0.005$ dex rms, or $1.2\%$, over the energy range $1-10$ GeV. 
We adopt $\delta\alpha = 0.15$ and $q_{\rm cal} = 0.77$\footnote{Neither of these sets the scale of our predictions. That is fixed by $\eta_c$: the disc luminosity is $\propto \eta_c f_{\rm cal}$ and the power delivered to the CGM $\propto \eta_c (1 - f_{\rm cal})$, so $q_{\rm cal}$ and $\delta\alpha$ only apportion a fixed budget between the two. The $3\%$ shift changes the disc contribution to the IGRB by $2.6\%$, from $0.62$ to $0.61$ of the background, and the power injected into the CGM by  $\leq4\%$
.}. 
We plot the comparison, overlaid with the observed IGRB, in \autoref{fig:IGRBspecFit}. The resulting calibrated $f_{\rm cal}$ as a function of present-day halo mass and redshift is plotted in \autoref{fig:plotfcal}.

\clearpage

\label{lastpage}
\end{document}